\documentclass[aps,prd,twocolumn,nofootinbib,nobibnotes]{revtex4-2}

\usepackage{physics}
\usepackage{graphicx}% Include figure files
\usepackage{dcolumn}% Align table columns on decimal point
\usepackage{bm}% bold math
\usepackage{hyperref}% add hypertext capabilities
\usepackage{xcolor}
\usepackage{cancel}
\usepackage{amsmath}
\usepackage[normalem]{ulem}
\usepackage{soul}
\usepackage{mathtools,slashed}

\newcommand{\preprintnumber}{\textbf{IPARCOS-UCM-25-045, ANL-200237}}

\newcommand{\ketbrac}[2] { {\text{$\mathfrak{C}_{{#1}{#2}}$}} }
\newcommand{\projector}[2]{{\text{$\mathbb{P}^{({#1})}_{{#2}}$}}}
\newcommand{\set}[1]{{\text{$\mathfrak{s}^{\dagger}_{{#1}}$}}}
\newcommand{\scrap}[1]{{\text{$\mathfrak{s}_{{#1}}$}}}

\begin{document}
% Hack: insert preprint number in top-right even in two-column mode
  \vspace*{-1.5cm}
 \hfill \preprintnumber
  \vspace*{1.2cm}
% END hack

%%%%%%%%%%%%%%%%%%%%%%%%%%%%%%%%%%%%%%%%
\title{\textbf{Quantum Computing Hadron Fragmentation Functions 
\\ in Light-Front Chromodynamics } 
}
%%%%%%%%%%%%%%%%%%%%%%%%%%%%%%%%%%%%%%%%%

\author{J. J. G\'alvez-Viruet, Felipe J. Llanes-Estrada, and Nicol\'as Mart\'{\i}nez de Arenaza}
 \affiliation{Depto. F\'{\i}sica Te\'orica \& IPARCOS, Univ. Complutense de Madrid, 28040 Madrid, Spain}

\author{Mar\'{\i}a G\'omez-Rocha}%
\affiliation{%
Depto. de F\'{\i}sica At\'omica, Molecular y Nuclear
and Instituto Carlos I de F\'{\i}sica Te\'orica y Computacional,
Universidad de Granada, 18071 Granada, Spain
}%

\author{T.~J.~Hobbs}
\affiliation{High Energy Physics Division, Argonne National Laboratory, Lemont, IL 60439, USA}

\date{\today}% It is always \today, today,

\begin{abstract}
We deploy Quantum Chromodynamics (QCD) in Light-front Quantization (and Gauge), discretized and truncated in both Fock--and momentum--spaces with a particle-register encoding suited for quantum simulation; we show for the first time how to calculate fragmentation functions, a problem heretofore untractable in general from \emph{ab-initio} approaches. We provide a classical-simulator based proof-of-concept by computing the charm-to-charmonium fragmentation, $c\to J/\psi$, in a simplified setup, an interesting case where we can (reasonably) compare with the known 1993 perturbative evaluation (and a more recent one at NLO) within Nonrelativistic QCD. 

\end{abstract}

\maketitle

%%%%%%%%%%%%%%%%%%%%%%%%%%%%%%%%%%%%%%%%%%%%%%%%
\section{The difficulty of computing \\ 
parton $\to$ hadron fragmentation}
%%%%%%%%%%%%%%%%%%%%%%%%%%%%%%%%%%%%%%%%%%%%%%%%
%
A standard method to study the structure of matter in fundamental physics is to break objects apart in an effort to reveal their internal constituents. In QCD, quarks and gluons cannot be detected directly as asymptotic states, but are confined into mesons and baryons; the final states of particle physics experiments are therefore governed by fragmentation, in which partons emerge from hard-scattering interactions and recombine into final-state hadrons, along the way, fusing with partons from the vacuum. This inverted logic, whereby hadronic structure is informed through hadronization rather than direct scattering, is unique to the strong interaction and makes the structure of a wider array of hadrons accessible, including short-lived resonances which cannot be directly probed in typical scattering experiments. 

At the partonic level, the probability of finding a hadron, $h$, accompanied by other unmeasured particles, $X_{\text{out}}$, in a jet initiated by a given bare parton, $j$, and carrying a longitudinal momentum fraction, $z=p_h^+/p_j^+$,
is encoded in the (unpolarized) fragmentation function (FF)~\cite{Gronau:1973gc,Feynman:1973xc} $D^h_{j}(z)$, which at fixed scale~\cite{Collins:2023cuo} is
\begin{eqnarray}
    D_{j}^h(z) 
    \equiv \frac{\rm Tr_{s}  \rm Tr_{c} } {3\,N_{c,j}}\sum_X 
    \langle j,p_j | h_s,X_{\text{out}}\rangle 
    \langle h_s,X_{\text{out}} | j,p_j\rangle 
     \label{FuncionFragmentacion}
\end{eqnarray}
with $N_{c,j} $ the number of, and the trace taken over the parton colors, respectively.  The importance of these FFs lies in their universality, which rests upon collinear factorization theorems: independently of the external process, the production of a given hadron from a given parton will always entail a convolution of some hard, perturbatively computable kernel $C_l$ with the same FF $D_j^h(z)$.
For example, production in $e^-e^+$ annihilation via a parton of label $j$ would be
\begin{equation}
\frac{d\sigma|_{e^-e^+\to h(z)X}}{dz}\!\!  = \!\! 
    \sum_p\!\! \int_z^1\!\! \frac{dy}{y}  \frac{d\sigma|_{e^-e^+\to j\, (y)X}}{dy}D_j^h(z/y).
    \label{factorization}
\end{equation}

FFs carry nonperturbative information about the confinement of the parton $j$ in the hadron $h$, and though fit to experimental high-energy data~\cite{AbdulKhalek:2022laj}, to date  they generally remain uncalculated from first-principles. This comes about because Lattice Gauge Theory, the {\it ab initio} method of choice for nonperturbative computations in QCD, is a Euclidean space formulation, analytically continuing time $t\to x_0=it$. The field-theory-based expression wrapping Eq.~(\ref{FuncionFragmentacion}) usable in a spacetime lattice is~\cite{Collins:2023cuo} however
\begin{eqnarray} \label{Ddefinition}
D_{ j}^h(z) = \frac{z^{1-2\varepsilon}\rm Tr_D}{4} \sum_{X_{\rm out}}
\int \frac{dx^+}{2\pi} e^{ik^-x^+} \gamma^- \cdot \nonumber \\ 
\langle 0 | \psi_j\left( \frac{x}{2}\right) | h,X_{\text{out}}\rangle     \langle h,X_{\text{out}} | \bar{\psi}_j \left( -\frac{x}{2}\right) | 0\rangle \ .
\end{eqnarray}
Connecting  $-x/2$ and $x/2$ along light-front time (because of the integral over $x^+$) requires an evolution operator along a lightlike direction. This is not possible in a Euclidean formulation in which all directions are spacelike; as such, the predominant approach has been to extract fragmentation functions through global phenomenological fits, {\it e.g.}, \cite{Bertone:2018ecm}. 
Quantum computers may change this situation by allowing direct computation of FFs as illustrated for the Nambu-Jona-Lasinio model~\cite{Li:2024nod}, or within the more fundamental Light-Front QCD (LFQCD) Hamiltonian as we demonstrate.

%%%%%%%%%%%%%%%%%%%%%%%%%%%%%%%%%%%%%%%%%%%%%%%%
\section{LFQCD on a quantum computer} \label{sec:H}
%%%%%%%%%%%%%%%%%%%%%%%%%%%%%%%%%%%%%%%%%%%%%%%%
For practical real-time simulations, we require an evolution operator which we can implement on quantum hardware or classical emulation platforms.
To this end, we use the light-front-time evolution operator of QCD in light-front gauge ($A^+\!=\!0$), $U(x^+)=e^{-ix^+P^-}$; we formulate $P^-$ in terms of the non-Abelian part of the chromomagnetic field, $\tilde{B}_{r, \mu \nu} = \sum_{s,t=1}^8 f^{rst} \tilde{A}_{r,\mu} \tilde{A}_{r,\nu}$, and color current,
$ \tilde{J}^\mu_{r} = \overline{\widetilde{\Psi}} \gamma^\mu T_r \widetilde{\Psi}+ \sum_{s,t=1}^8 f^{rst} (\partial^\mu \tilde{A}^\nu_{s}) \tilde{A}_{t,\nu} $ ($r,s=1\dots 8$). We therefore implement the light-front Hamiltonian in standard notation~\cite{Brodsky:1997de},
\begin{flalign}\label{hamiltoniano}
& P^- = %\quad 
\int dx_+d^2x_\perp \Bigg\{ \frac{1}{2}\left( 
\overline{\widetilde{\Psi}} \gamma^+ \frac{m^2 - \nabla^2_\perp }{i \partial^+}\widetilde{\Psi} 
- \widetilde{A}_r^\mu \nabla^2_\perp \widetilde{A}_{r,\mu } \right) \nonumber 
\\ &\quad 
+ g\, \widetilde{J}^\mu_r \widetilde{A}_{r,\mu } 
+ \frac{g^2}{4} \, \widetilde{B}^{\mu \nu}_r \widetilde{B}_{r, \mu \nu}\, -\, \frac{g^2}{2} \, \widetilde{J}^\mu_r \frac{1}{(\partial^+)^2} \widetilde{J}^\mu_r
\nonumber 
\\  &\quad 
- \frac{g^2}{2} \, i \overline{\widetilde{\Psi}} \gamma^\mu T_r \widetilde{A}_{r,\mu} 
\frac{\gamma^+}{\partial^+} (\gamma^\nu T_s \widetilde{A}_{s,\nu} \widetilde{\Psi}) \Bigg\}\ .
\end{flalign}
We proceed by expressing~(\ref{hamiltoniano}) in terms of normal modes in momentum space, quarks ($b^\dagger$), antiquarks ($d^\dagger$), and gluons ($a^\dagger$), satisfying canonical commutation relations:
\begin{multline} \label{canonical}
\left[a_p, a^\dagger_{k} \right] = \left\{b_p, b^\dagger_{k} \right\} =\left\{d_p, d^\dagger_{k} \right\}=\delta_{p,k}\\ \equiv \delta(p^+-k^{+})\delta^2\left(p_\perp-k_\perp\right) \delta^{\sigma '}_{\sigma}\delta^{r'}_{r}\ ,
\end{multline}
where we adopt the abbreviation $a_{p\sigma r}=a_p$. For example, the (transverse) kinetic-energy operator is, in terms of parton auxiliary kinetic functions $t_i$, 
\begin{eqnarray} \label{kinetic}
    E_c=\sum_p \left[t_q(p)b^\dagger_pb_ p +t_{\overline{q}}(p)d^\dagger_pd_ p+t_g(p) a^\dagger_p a_ p    \right],  \\
    t_q(p)\equiv t_{\overline{q}}(p)\equiv \left(\frac{m^2+p^2_\perp}{x_q} \right)_{q}\ ,\    t_g(p)\equiv \left(\frac{p^2_\perp}{x_g} \right)_g\ \nonumber.
\end{eqnarray}
The interaction terms in this Hamiltonian (field-theory potentials) are long and are instead provided in appendix \ref{app:H}. They are classified~\cite{Brodsky:1997de} as $H_I=V+F+S+C$ and named according to the respective diagrams to which they give rise: \textit{vertices} ($V$), like the gluon bremsstrahlung terms, $a^\dagger b^\dagger b $; \textit{forks} ($F$), such as $ b^\dagger d^\dagger d^\dagger d $, where the instantaneous light-front Coulomb-like potential emits a pair; \textit{seagulls} ($S$), such as the non-Abelian coupling, $a^\dagger a^\dagger aa$;  and \textit{contracted potentials} ($C$) that yield one-body self-energies or vacuum polarizations, {\it e.g.}, $b^\dagger b$.

Discretization forces momentum cutoffs, $\lambda\! <\! p\! <\! \Lambda$, and a maximum number of particles, $N_{\text{max}}$; in this demonstration, we truncate at two quarks, one antiquark, and one gluon --- limiting the terms of $H$ we can implement. We expect progress in the field will bring the dependence on these two parameters under control with Wilsonian-type or other renormalization-group equations~\cite{Gomez-Rocha:2015esa}.

We have developed an encoding~\cite{Galvez-Viruet:2024hry}  (see App.~\ref{app:encoding}), that allows us to program products of these particle creation/destruction operators with arbitrary (discretized) functions of momenta.
The basic ingredients are, first, the ``particle register'', a memory unit that contains a presence/absence qubit tracking whether the register is active and typically used as a gate control, and qubits for color, spin/polarization, momentum and quark flavor; second, 
set/scrap operators, $\mathfrak{s}$, $\mathfrak{s}^\dagger$, which alter the quantum numbers in that register by writing or erasing as needed; third, symmetrizer, $\mathcal{S}$, and antisymmetrizer, $\mathcal{A}$, operators, which are applied every time a particle is created or erased from memory and which guarantee the correct statistics for each particle species. 
Lie exponentials of the various terms in $P^-$ [Eq.~(\ref{hamiltoniano})] which combine these ingredients have been coded in {\tt Python} and IBM's {\tt Qiskit} package~\cite{Javadi-Abhari:2024kbf} and simulated with the {\tt statevector} backend. We present a first exploratory application to a nontrivial physical problem; other colleagues have employed related methods to assess {\it in medio} jet-energy loss~\cite{Barata:2023clv}.

\begin{figure}[h]
\includegraphics[width=0.9\columnwidth]{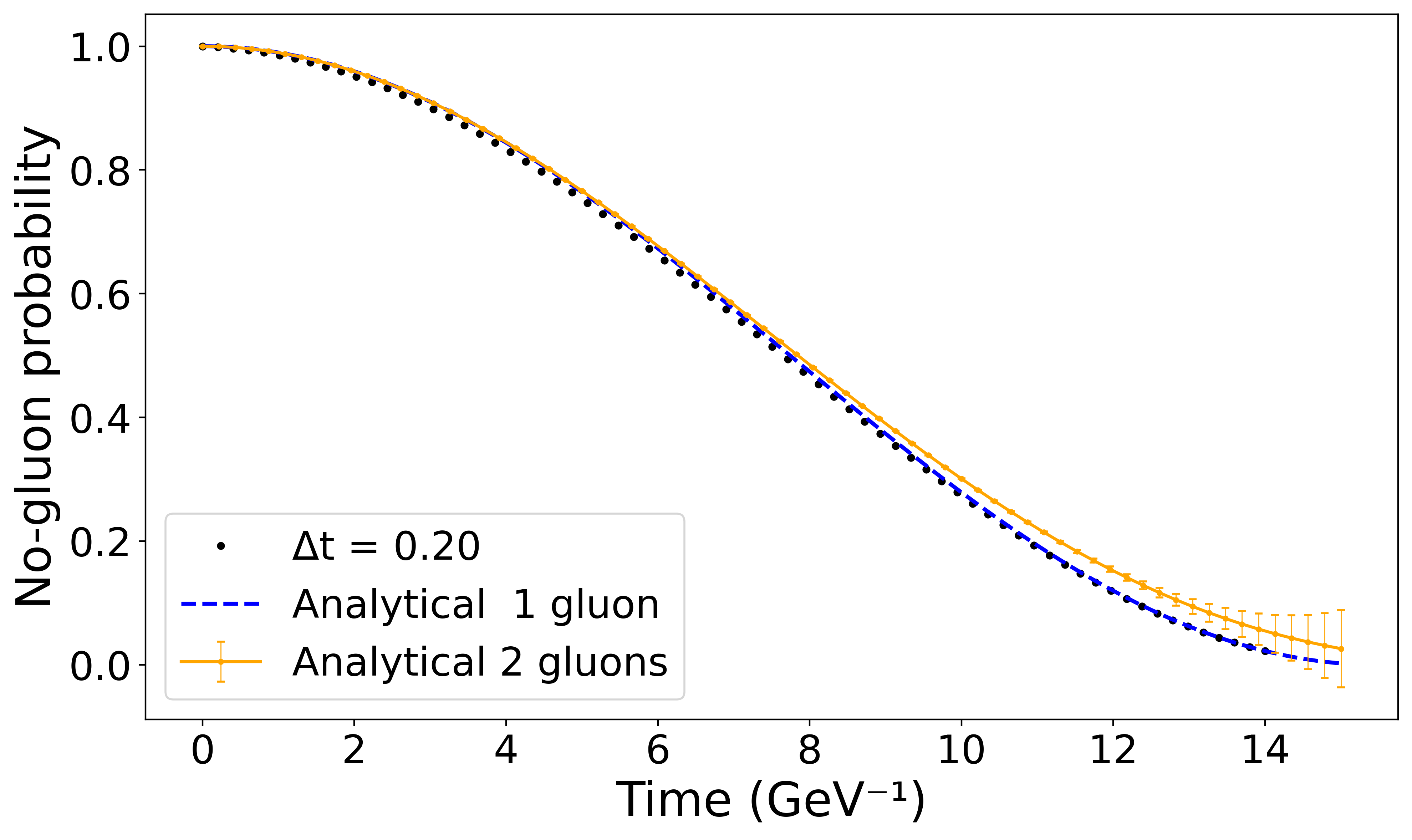}
\includegraphics[width=0.9\columnwidth]{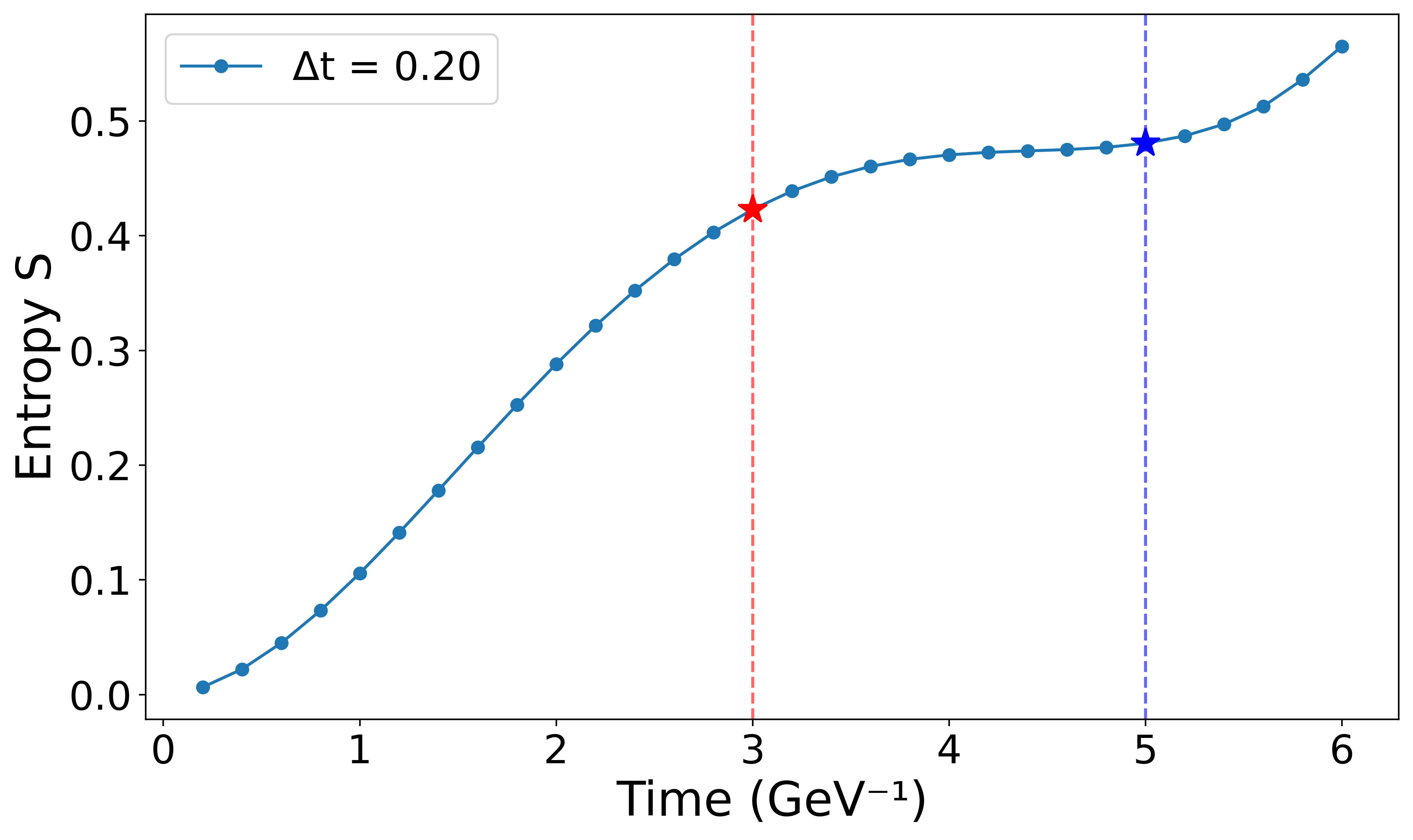}
\caption{\label{fig:Entropy} $SU(2)$ simulation with up to four particles in an ($N=8$) $k^+$ grid. Upper plot: the probability ({\it vs}.~time) for an initial $c$-quark with maximum grid momentum to not be accompanied by a bremsstrahlung gluon for $H$ restricted to $V_1$ interactions with a simulated quantum memory able to hold 1 or 2 gluons. Lower plot: The entropy, $S(t)$, increases with the average number of modes populated by radiation from the initial parton for $H$ restricted to the kinetic terms and $V$ interactions. Ideally, $D_j^h$ is to be extracted around saturation (maximum entropy); given that errors (upper plot) increase with time, we extract it  when a plateau is found (between the red and blue vertical lines).}

\end{figure}

Discretization of the longitudinal momentum fraction $z$ with a minimum $z_{\rm min} = 1/N$ value entails that the ``zero modes'' of any fields, having $p_h^+=0$ or $p_j^+=0$ are not represented in the computation from the start. These modes are not directly accessible to experiment, either in the initial or final states, through parton distribution or fragmentation functions, due to the finite momentum resolution inherent to any measurement.

Given that in LF--quantization modes with $p^+>0$ cannot condense in the vacuum, the exclusion of zero modes leaves out the physics of dynamical chiral symmetry breaking~\cite{Kim:1994rm,Brodsky:2012ku,Tsujimaru:1997jt}. However, this is a prominent feature of the low-lying hadron spectrum which is less important at high excitation~\cite{Bicudo:2016eeu} and seldom of interest at the highest energies of jet physics and fragmentation functions. 
Should zero modes be incorporated in future computations along the lines here proposed, care should be taken to avoid gauge redundancies, as the $A^+=0$ gauge-fixing tolerates residual gauge transformations. This is because
\begin{equation}
\delta A^{+a} = \partial^+ \omega +g f^{abc} A^{+b} \omega^c  
\end{equation}
vanishes (and thus respects the gauge condition $A^+=0$) as long as $\partial^+ \omega^a = 2\frac{\partial}{\partial x^-} \omega^a=0$, and the resulting  
$\omega^a=\omega^a(x^+,x^\perp)$ is precisely a zero mode, because of its $x^-$--independence.
The redundancy would be solved by imposing the zero-mode part of Gauss's law on the physical states, 
\begin{eqnarray}
Q^a(x^+)|{\rm phys}\rangle =0\nonumber \\
Q^a = \int dx^- \int d^2 x_\perp J^{+a} (x)\ .
\end{eqnarray}
We do not need to explore these operators at present, as the zero modes currently fall out of the  implemented grid.
\color{black}
%%%%%%%%%%%%%%%%%%%%%%%%%%%%%%%%%%%%%%%%%%%%%%%%
\section{Simulating fragmentation functions}
%%%%%%%%%%%%%%%%%%%%%%%%%%%%%%%%%%%%%%%%%%%%%%%%
Contemporary quantum hardware (as of 2025-26) produce high noise levels for computations like those outlined above, given the low fidelity (98-99.5\%) of current two-qubit gates and fast decoherence times (qubit relaxation and dephasing due to the environment) of order 100-400$\mu$s; these limitations constrain actual circuits to a few tens of layers deep at best, but not the at least $10^7$--$10^8$  gates which these types of simulations require (see App. \ref{sec:scaling}), similarly to those in nuclear physics~\cite{Carrasco-Codina:2025gan}. Our results are therefore computed on classical simulators. Because the memory requirements to represent an $N$-qubit Hilbert space exponentiate, we are limited to the equivalent of about 30 qubits. Still, these modest simulations offer quite some insight in how upcoming quantum computers shall behave and what kind of results they can produce. 

Unitary evolution on a quantum computer can be decomposed using a Lee-Trotter expansion, $U( \Delta t)=e^{-iT \Delta t} \prod_j  e ^ {-iH_{Ij} \Delta t} + O[(\Delta t)^2]$. As successive rotations $U_i$ are applied to a discretized system $| \psi_0\rangle $ with a finite number of modes, it can cycle back to something very close to the original state (in the upper plot of Figure~\ref{fig:Entropy} the continuous line is $\cos^2(Vx_+)$ with $V$ the relevant potential to emit a gluon from the initial jet $c$-quark, and the dots the simulation outcomes). 

Therefore, a criterion is needed to stop the evolution and extract $D_j^h$ at the corresponding light-front time. We adopt a condition based on the probability spreading among the many degrees of freedom (as many quarks and gluons as are compatible with the given energy), increasing the Shannon entropy, $S$, over all the modes active in the computer memory (distinct from the entropy of the fragmentation function or related objects~\cite{Benito-Calvino:2022kqa,Bloss:2025ywh}). 

A small system under unitary evolution cycles back, as in the upper panel of Figure~\ref{fig:Entropy}.  With enough parton emission, however, $S$ increases until the computer memory is saturated and no further significant evolution occurs (maximum entropy): then we can consider that the limit $x^+\!\to\! \infty$ has been reached for any practical purpose, and accordingly stop the evolution to extract $D_j^h(z)$. For the time being, (lower plot of Figure~\ref{fig:Entropy}) we are content with finding a plateau of $S$ before the uncertainties due to Fock-space truncation become important.
Obviously, the use of entropy is a poor man’s device to track whether the available number of particles in the memory has been saturated. Ideally, one would run with a variable number of particles and wait until $D(z)$ appears to be evolution-time independent (and check the entropy too). But for the time being, we do not believe that much more can be done until actual quantum computers become practical, because classical simulators are memory-limited to undertake this physics.

We exemplify the procedure with the $J/\psi$, an ``easy'' quarkonium meson for which the quark-model representation with a spin-1 color-singlet quark-antiquark pair with respective momentum fractions $x<z$ and $z-x$,
\begin{eqnarray} \label{Jpsiwf}
|J/\Psi\rangle =  \sum \delta_{ c_qc_{\bar{q}} }  \frac{\chi_0(x) \,\vec{\sigma}_{ij}}{\sqrt{x \,(z-x)}}
\left|x\ i\,c_q ,(z-x)\ j\,c_{\bar{q}}\right\rangle \ , \ \ 
\end{eqnarray}
 is a reasonable first approximation. 
 In this calculation, we only treat longitudinal momentum and set $p_\perp\!=\!0$. Ideally, the longitudinal wavefunction should be estimated
by minimizing the Hamiltonian of Eq.~(\ref{hamiltoniano}), but use here a simple ansatz based upon a 1-D distribution~\cite{Li:2021cwv} with Cisneros' end-point suppression,
\begin{equation} \label{Cisneroschi}
\chi_0(x) = \frac{1}{\sqrt{C}} \  x^{\beta/2} (z - x)^{\alpha/2}\ ,
\end{equation}
where the normalization $C$ is chosen to obtain unit square norm over $x\in [0,z]$.
We take the parameters of this simple longitudinal wavefunction to be
$\alpha = 4 m_q^2/\kappa$, $\beta = 4 m_{\bar{q}}^2/\kappa$, with $m_q = m_{\bar{q}} = 1.27$ GeV and $\kappa = 1.34$ GeV.

\begin{figure}
\includegraphics[width=0.95\columnwidth]{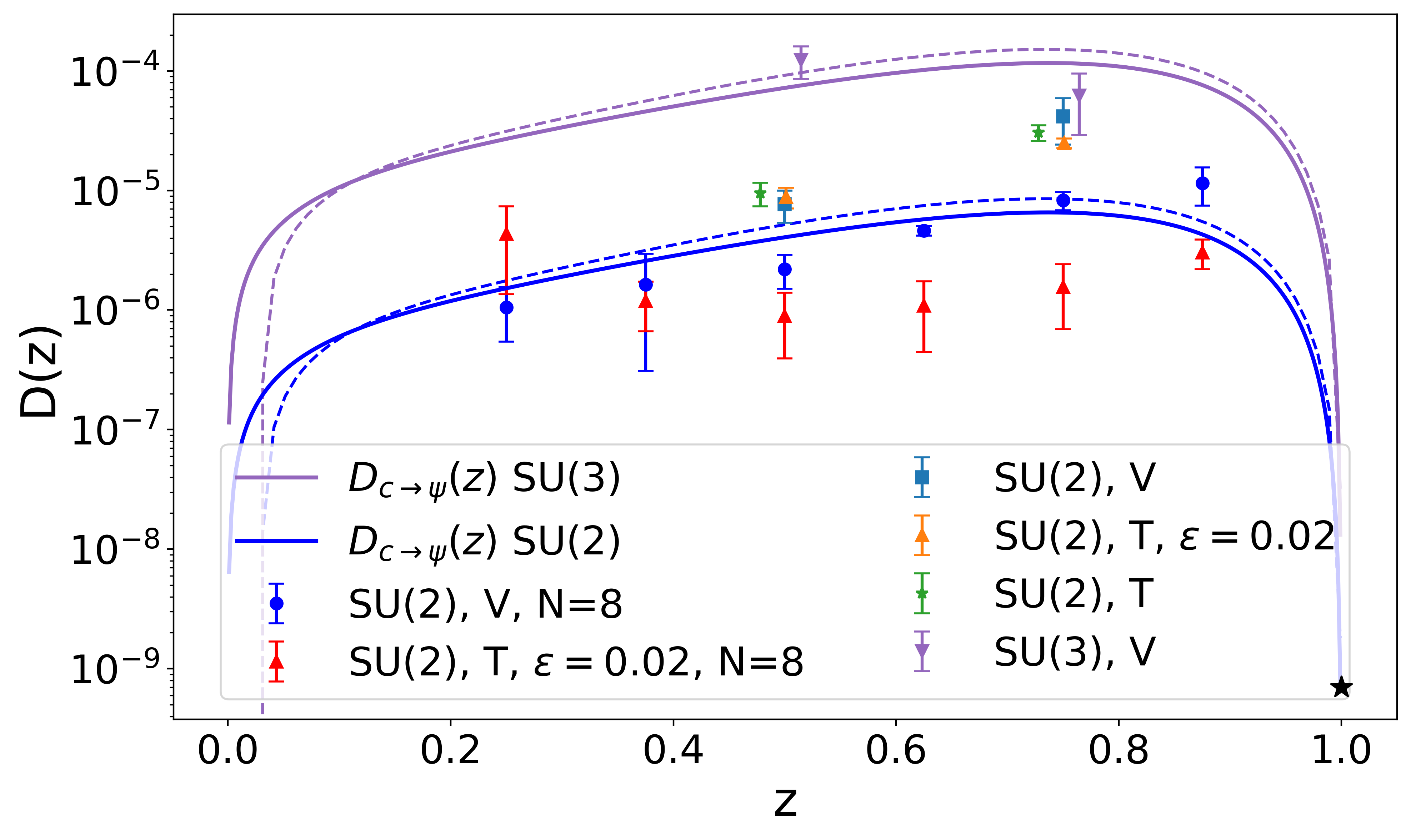}
\caption{\label{fig:JpsifragF} 
Fragmentation function of a $c$-quark to a $J/\psi$ meson. $T$ denotes the total Hamiltonian ($V$ the vertex-type terms only); the $k^+$ momentum lattice size is set to $N=4$ (or $N=8$ if indicated).
The reference lines are the NRQCD computation at LO~\cite{Braaten:1993mp} and NLO~\cite{Zheng:2019dfk} (dashed). 
The symbols stem from our classical simulation of what the quantum computer would find, as in Figure~\ref{fig:Entropy}, for various $J/\psi$ momentum fractions $z$. The uncertainty raisers span only the time interval for $D$ extraction (lower plot of Fig.~\ref{fig:Entropy}).
By construction of the numerical grid, the endpoint value $D(1)=0$   vanishes as it must.}
\end{figure}

Fig.~\ref{fig:JpsifragF} shows the extracted fragmentation function for that $J/\psi$ model at various times up to 1 GeV$^{-1}$, where entropy has already reached the plateau (probability spread among several modes). Likewise, the $J/\psi$ already had a reasonable time to form as the characteristic scale is the meson radius, estimated~\cite{Hatton:2020lnm} to be 
$0.321 \pm 0.014\ \text{fm}\simeq 1.63 \pm 0.07\ \text{GeV}^{-1}$.

We see that the uncertainty bands due to the choice of extraction time corresponding to the FF at various $z$ broadly agree with the NRQCD results~\cite{Braaten:1993mp,Zheng:2019dfk}, which is reassuring given the simplicity of this first demonstration. One cannot hope for more accuracy yet, given the cutoffs imposed both at the particle and at the momentum level, and a renormalization scheme has not been deployed (the fragmentation function must depend on that scheme, while its subsequent convolution with the hard kernel should not). Also the $J/\psi$ state should be extracted from minimization of $H_{\text{QCD}}$ in the same scheme, and Eq.~(\ref{Cisneroschi}) has to be seen as a variational approximation to that state. We leave these questions to future work.

To extract the numerical data in Fig.~\ref{fig:JpsifragF} after the light-front time evolution, we need to measure whether a $J/\psi $ meson has been fragmented. For this we introduce the annihilation gate which takes the meson to the vacuum, in analogy to an interpolating operator in lattice theory,
\begin{equation}
    U_{\overline{Q}Q}(k) |\psi^{m}_{J/\Psi}(k)\rangle = |\widetilde{\Omega}\rangle_q  |\widetilde{\Omega}\rangle_{\overline{q}}
\end{equation}
Thus, after applying this $U_{\overline{Q}Q}$ gate, if we measure the $q\overline{q}$ registers and find them empty, it will mean that the $J/\psi$ was present at the end of the evolution.

%%%%%%%%%%%%%%%%%%%%%%%%%%%%%%%%%%%%%%%%%%%%%%%%
\section{Computational scaling and Outlook}
%%%%%%%%%%%%%%%%%%%%%%%%%%%%%%%%%%%%%%%%%%%%%%%%

In this calculation we have employed a memory which can hold two quarks (with spin, color, flavour), an antiquark, and a gluon (with one more qubit for color, but no flavor), with 2 and 3 longitudinal--momentum qubits respectively, in both $SU(2)$ (total of 21 qubits) and $SU(3)$ (25 qubits). With four additional ancillary qubits, the total counts become 25 and 29, respectively. With a few hundred well-working qubits the calculation could be extended to a serious number of particles and reach lower $x$ by expanding the momentum grid, well beyond the  valence--like regime explored here. One more order of magnitude to the $O(10^3)$ qubits then allows to explore the $p_\perp$ distribution and perform proper extrapolations in the cutoffs and the Trotter time step. This large number of qubits is currently inaccessible to classical simulators, and less noisy quantum computers are needed.

A short discussion about the uncertainties affecting this work is in order. The first consideration is the effect of truncating the Fock space to at most four particles ($2\times q+\overline{q}+g$). While we cannot estimate the uncertainty brought about by the large number of Hamiltonian terms which link this state to higher Fock sectors, we can and have gotten an idea by analyzing additional insertions of $V_1$ (the gluon emission and absorption vertex) as discussed in appendix~\ref{app:uncertainty}.  

Further, the discretization of the longitudinal momentum, $P^+$, deserves our attention. Employing only three qubits, to yield an 8-state Hilbert space for each momentum, means that the emitted gluon has little phase space and some vertices like $V_2$ which otherwise are nonvanishing matrices in the four-particle subspace are forced to vanish. The absence of $P_\perp$ in the description of the bound state being produced, on the other hand, is expected to induce errors of order $\text{max}(P_\perp)/\text{min}(P_+)$.
Finally, the finite time step induces a further discretization error which can be read off Figure~\ref{fig:Entropy} near the saturation of the entropy, to be of order 8-10\%.
Further controlling these sources of uncertainty will be essentail to the continued development of light-front quantum simulation.

As phenomenological fits of fragmentation functions have generally concentrated on the light-quark sector,
charm hadronization has only been calculated within effective formalisms
like NRQCD; we thus compare our results against representative such calculations~\cite{Braaten:1993mp,Zheng:2019dfk}.

As for quantum hardware, the gate--implementation time has considerably diminished 
from around 130 ns in 2017 to 60--88 ns in 2025~\cite{Linke:2017knx}, with the number of qubits exponentiating with Rose's Moore-like law and currently at $O(10^2-10^3)$.  This will soon suffice for the calculations seemingly needed to carry the program here proposed. However, significant improvements are necessary in the error rate (which should drop from percent-level to the range $10^{-8}$, and decoherence time (which should increase from $O(10^{-4}-10^{-3})$ to $O(10-100)$ seconds.   

In summary, we believe that we have demonstrated a beginning-to-end method to extract fragmentation functions which will become very attractive with quantum hardware advancements.

%%%%%%%%%%%%%%%%%%%%%%%%%%%%%%%%%%%%%%%%%%%%%%%%

%%%%%%%%%%%%%%%%%%%%%%%%%%%%%%%%%%%%%%%%%%%

%%%%%%%%%%%%%%%%%%%%%%%%%%%%%%%%%%%%%%%%%%%%%%%%
\begin{acknowledgments}
%%%%%%%%%%%%%%%%%%%%%%%%%%%%%%%%%%%%%%%%%%%%%%%%
We thank A.~H.~de Tejada Gross and D. Fern\'andez Sanz for assistance, and S.J. Brodsky as well as M.P.~Zurita for encouragement and useful comments.
Work partially supported by grants 
PID2023-147072NB-I00; %<--Maria
PID2022-137003NB-I00 %<--Complutense
and FPU21/04180 %<-- JJ scholarship
of the Spanish MCIN/AEI/10.13039/501100011033/ and Ministry of Universities;
``Ayudas de Máster IPARCOS-UCM/2024'';  %<-- Nicolas scholarship
and the  U.S. Department of Energy under contract DE-AC02-06CH11357.
Numerical computing carried out at Granada's PROTEUS supercomputer.

%%%%%%%%%%%%%%%%%%%%%%%%%%%%%%%%%%%%%%%%%%%%%%%%
\end{acknowledgments}
%%%%%%%%%%%%%%%%%%%%%%%%%%%%%%%%%%%%%%%%%%%%%%%%

\appendix

%%%%%%%%%%%%%%%%%%%%%%%%%%%%%%%%%%%%%%%%%%%%%%%
\section{General conventions} \label{apendice}
%%%%%%%%%%%%%%%%%%%%%%%%%%%%%%%%%%%%%%%%%%%%%%%
\paragraph{Dirac spinors} $u_\alpha(p,\lambda)$ and $v_\alpha(p,\lambda)$ are solutions of Dirac's equation that, written in terms of Dirac's gamma matrices
$\cancel{p}=p_\mu\gamma^\mu$, is
\begin{multline}
    \left(\cancel{p}-m \right)u_\alpha(p,\lambda)=0,\enskip \left(\cancel{p}+m \right)v_\alpha(p,\lambda)=0\ . 
\end{multline}
Additionally, the spinors form an orthogonal basis. Explicitly,
\begin{align}
    \overline{u}_\alpha(p,\lambda)\, u_\alpha(p,\lambda') &= \overline{v}_\alpha(p,\lambda')\, v_\alpha(p,\lambda) = 2m\, \delta_{\lambda\lambda'} \nonumber\\
    \sum_\lambda u_\alpha(p,\lambda)\, \overline{u}_\alpha(p,\lambda) &= \slashed{p} + m \nonumber \\
    \sum_\lambda v_\alpha(p,\lambda)\, \overline{v}_\alpha(p,\lambda) &= \slashed{p} - m.
\end{align}

%%%%%%%%%%%%%%%%%%%%%%%%%%%%%%%%%%%%%%%%%%%%%%%
\paragraph{Gluon polarization vectors} 
The four-vectors $\epsilon_\mu(p,\lambda)$ contain two physical polarizations
which can be taken to be the  projections of a spin--1 ordinary vector over the axis $OZ$, tagged by $\lambda=\pm 1$, 
\begin{equation}
    \epsilon_\perp \left( \lambda \right)=\frac{-1}{\sqrt{2}}\left( \lambda e_x+ ie_y\right),
\end{equation}
with $e_x$ and $e_y$ of unit norm. Because $\epsilon_+ \left(\lambda\right)=0$  due to the light--front gauge choice $A^+=0$, the polarization four--vector can be written as
\begin{equation}
    \epsilon_\mu\left(p,\lambda\right)=\left(0,\frac{2\epsilon_\perp\left(\lambda \right)p_\perp}{p^+}, \epsilon_\perp\left(\lambda\right)\right). 
\end{equation}

They are also an orthogonal basis and because they are transverse, they generate the linear space of solutions to the free Maxwell's equations in this gauge.
The orthogonality, transversality and closure relations read
\begin{multline}
    \epsilon^\mu(p,\lambda)\epsilon_\mu^\star(p,\lambda')=-\delta_{\lambda,\lambda'}, \enskip p_\mu \epsilon_\mu (p,\lambda)=0, \\     \sum_\lambda\epsilon_\mu(p,\lambda)\epsilon_\nu^\star(p,\lambda)=-g_{\mu\nu}+\frac{\eta_\mu p_\nu+\eta_\nu p_\mu}{p^k\eta_k}, 
\end{multline}
(with an auxiliary null vector $n^\mu\equiv \left(0,2,0^{\perp}\right)$ being used in the last line).

Because the strong interaction is flavor--independent, the flavor indices in all the tables which follow will appear only in Kronecker deltas $\delta_{l_1}^{l_2}$.

In terms of these spinors and polarization vectors we can now quote all terms of the Hamiltonian of QCD in the light front, which we spell out in the next section.
%%%%%%%%%%%%%%%%%%%%%%%%%%%%%%%%%%%%%
\section{Complete light--front gauge QCD  Hamiltonian}\label{app:H}
%%%%%%%%%%%%%%%%%%%%%%%%%%%%%%%%%%%%%
The Hamiltonian naturally splits into non-interacting and interacting parts, $H=E_c+H_I$. 
The normal--mode expansion of $E_c$ has already been given, to exemplify, in Eq.~(\ref{kinetic}). Here we will briefly recall the interacting terms. Two of us (N. Mart\'{\i}nez de Arenaza and M. G\'omez Rocha) have completely recalculated the Hamiltonian. 

We use a compact notation in which we employ only one subindex, {\it e.g.} $b^\dagger_p$ (see for example Eq.~(\ref{canonical}) in the main text). Typically we choose either a numeral to select which of several quantum--number sets one is dealing with, or simply the momentum subindex $p$, in the understanding that other quantum numbers (color, flavor, spin or polarization) are represented by that index too.

%%%%%%%%%%%%%%%%%%%%%%%%%%%%%%%%%%%%%
\subsection{Vertex (splitting) interactions}
%%%%%%%%%%%%%%%%%%%%%%%%%%%%%%%%%%%%%%%%%%%%
First, we describe the ``Vertex'' $V$ terms which provides the core of any fragmentation--function study initiated from a quark.

The gluon emission or absorption vertex $V$ is
\begin{multline}
    V=\sum_{1,2,3} \Bigr[b^\dagger_1b_2a_3V_1\left(1;2,3 \right)  +  d^\dagger_1d_2a_3V_2\left(1;2,3 \right) \\+ a^\dagger_1d_2b_3 V_3\left(1;2,3 \right) 
    + a^\dagger_1a_2a_3V_4\left(1;2,3 \right) + h.c\Bigr] . \label{V_terminos_ec}
\end{multline}

Here, the functions $V_n(1;2,3)$ depend on the momentum variables $ q^+$, $q_\perp$; on color, flavor and helicity; and are given in tables   \ref{tab:V} and \ref{tab:V_x}. The matrix elements from table~\ref{tab:Miguales} have been employed in calculating them.
The compact notation employed therein shortens the subindices such as $1,2$ to mean $p_1, p_2$,  $b_n\equiv b(p_n)$ and $V_n(1;2,3)\equiv V_n(p_1;p_2,p_3)$. 
In consequence, $\sum_{1,2,3}$ adds up the three discretized momenta.
Moreover, $V_2\equiv -V_1^\star$. At last, we have employed the notation
 $V_{i}\equiv\sum_j V_{i,j}$; this has also been employed for the remaining terms of the Hamiltonian listed in the next subsection, of the type $F$, $S$ and $C$.

\begin{table}
    \centering
    \caption{
    Interaction vertices of type ``V'' in terms of Dirac spinors and gluon polarizations. Here, $\bigtriangleup v= \frac{g P^+ }{\sqrt{2\left(2 \pi \right)^3}} \delta\left(q^+_1-q^+_2-q^+_3 \right)$ $\delta^{(2)}\left(\Vec{q}_{\perp ,1}-\Vec{q}_{\perp ,2}-\Vec{q}_{\perp ,3} \right).$
 }
    \label{tab:V}
    \begin{tabular}{|c|c|}
    \hline
        $V_1$ &  $\frac{\bigtriangleup v}{\sqrt{q^+_1q^+_2q^+_3}} \left(  \overline{u}_1  {\cancel{\epsilon}_3} T^{a_3}u_2  \right)$ \\     \hline
        $V_2$ &  -$\frac{\bigtriangleup v}{\sqrt{q^+_1q^+_2q^+_3}} \left(  \overline{v}_2  {\cancel{\epsilon}_3} T^{a_3}v_1  \right)$ \\     \hline
         $V_3$&  $\frac{\bigtriangleup v}{\sqrt{q^+_1q^+_2q^+_3}} \left(  \overline{v}_2  {\cancel{\epsilon}^\star_1} T^{a_1}u_3  \right)$\\     \hline
         $V_4$& $\frac{f_{a_2a_3}^{a_1} \bigtriangleup v}{\sqrt{q^+_1q^+_2q^+_3}} \left[ \left(\epsilon_1^\star q_3 \right)\left(\epsilon_2 \epsilon_3 \right) + \left(\epsilon_3 q_1 \right)\left(\epsilon_1^\star \epsilon_2 \right)+\left(\epsilon_3 q_2 \right)\left(\epsilon_1^\star \epsilon_2 \right)  \right]$\\     \hline
    \end{tabular}

\end{table}

\begin{table}
    \centering
        \caption{List of necessary $\overline{u}(p) M u(q)$ and  $\overline{v}(p) M u(q)$ Matrix elements of each of the listed operators, generically $M$, in the Dirac-spinor basis.}
    \label{tab:Miguales}
    \resizebox{6.0cm}{!} {
    \begin{tabular}{|c|c|}
    \hline
        $M$ & $\frac{1}{\sqrt{p^+q^+}} \left(\overline{u}(p) M u(q)\right)\delta_{\lambda_p,\lambda_q}$ \\ \hline
        $\gamma_\perp \cdot a_\perp$ & $a_\perp \cdot \left(\frac{p_\perp}{p^+} + \frac{q_\perp}{q^+}\right) - i\lambda_q a_\perp \wedge \left(\frac{p_\perp}{p^+} - \frac{q_\perp}{q^+}\right)$ \\ \hline
        $\gamma^+$ & $0$ \\ \hline
        $M$ & $\frac{1}{\sqrt{p^+q^+}} \left(\overline{u}(p) M u(q)\right)\delta_{\lambda_p,-\lambda_q}$ \\ \hline
        $\gamma_\perp \cdot a_\perp$ & $a_\perp(\lambda_q) \left(\frac{m}{p^+} - \frac{m}{q^+}\right)$ \\ \hline
        $\gamma^+$ & $2$ \\ \hline
        $M$ & $\frac{1}{\sqrt{p^+q^+}} \left(\overline{v}(p) M u(q)\right)\delta_{\lambda_p,\lambda_q}$ \\ \hline
        $\gamma_\perp \cdot a_\perp$ & $a_\perp(\lambda_q) \left(\frac{m}{p^+} + \frac{m}{q^+}\right)$ \\ \hline
        $\gamma^+$ & $0$ \\ \hline
        $M$ & $\frac{1}{\sqrt{p^+q^+}} \left(\overline{v}(p) M u(q)\right)\delta_{\lambda_p,-\lambda_q}$ \\ \hline
        $\gamma_\perp \cdot a_\perp$ & $a_\perp \cdot \left(\frac{p_\perp}{p^+} + \frac{q_\perp}{q^+}\right) - i\lambda_q a_\perp \wedge \left(\frac{p_\perp}{p^+} - \frac{q_\perp}{q^+}\right)$ \\ \hline
        $\gamma^+$ & $2$ \\ \hline
    \end{tabular}
    }
\end{table}

\begin{table}[h!]
    \centering
    \caption{Interaction vertices in terms of helicity and the color factor, with   $\bigtriangleup v$ given in table~\ref{tab:V}.}
    \label{tab:V_x}    
    \begin{tabular}{|c|c|}

    \hline
        $V_{1.1}$ &  $\frac{\bigtriangleup v}{\left(P^+\right)^{3/2}}\sqrt{\frac{1}{x_3}}m_F\left[\frac{1}{x_1}-\frac{1}{x_2}\right]\delta^{\lambda_2}_{-\lambda_1}\delta^{\lambda_3}_{\lambda_1}\enskip \delta^{f_2}_{f_1}\enskip T^{a_3}_{c_1c_2}$ \\   
        \hline
         $V_{1.2}$ &  $\frac{\bigtriangleup v}{\left(P^+\right)^{3/2}}\sqrt{\frac{2}{x_3}}\Vec{\epsilon}_{\perp,3}\cdot \left[\left(\frac{\Vec{q}_{\perp}}{x}\right)_3-\left(\frac{\Vec{q}_{\perp}}{x}\right)_2\right]\delta^{\lambda_2}_{\lambda_1}\delta^{\lambda_3}_{\lambda_1}\enskip \delta^{f_2}_{f_1}\enskip T^{a_3}_{c_1c_2}$ \\   
        \hline
         $V_{1.3}$ &  $\frac{\bigtriangleup v}{\left(P^+\right)^{3/2}}\sqrt{\frac{2}{x_3}}\Vec{\epsilon}_{\perp,3}\cdot \left[\left(\frac{\Vec{q}_{\perp}}{x}\right)_3-\left(\frac{\Vec{q}_{\perp}}{x}\right)_1\right]\delta^{\lambda_2}_{\lambda_1}\delta^{\lambda_3}_{-\lambda_1}\enskip \delta^{f_2}_{f_1}\enskip T^{a_3}_{c_1c_2}$ \\   
        \hline
        $V_{3.1}$ &  $\frac{\bigtriangleup v}{\left(P^+\right)^{3/2}}\sqrt{\frac{1}{x_1}}m_F\left[\frac{1}{x_2}+\frac{1}{x_3}\right]\delta^{\lambda_3}_{\lambda_2}\delta^{\lambda_3}_{\lambda_1}\enskip \delta^{f_3}_{f_2}\enskip T^{a_1}_{c_2c_3}$ \\   
        \hline
         $V_{3.2}$ &  $\frac{\bigtriangleup v}{\left(P^+\right)^{3/2}}\sqrt{\frac{2}{x_1}}\Vec{\epsilon}_{\perp,1}\cdot \left[\left(\frac{\Vec{q}_{\perp}}{x}\right)_1-\left(\frac{\Vec{q}_{\perp}}{x}\right)_3\right]\delta^{\lambda_3}_{-\lambda_2}\delta^{\lambda_3}_{-\lambda_1}\enskip \delta^{f_3}_{f_2}\enskip T^{a_1}_{c_2c_3}$ \\   
        \hline
        $V_{3.3}$ &  $\frac{\bigtriangleup v}{\left(P^+\right)^{3/2}}\sqrt{\frac{2}{x_1}}\Vec{\epsilon}_{\perp,1}\cdot \left[\left(\frac{\Vec{q}_{\perp}}{x}\right)_1-\left(\frac{\Vec{q}_{\perp}}{x}\right)_2\right]\delta^{\lambda_3}_{-\lambda_2}\delta^{\lambda_3}_{\lambda_1}\enskip \delta^{f_3}_{f_2}\enskip T^{a_1}_{c_2c_3}$ \\    
        \hline
        $V_{4.1}$ &  $-\frac{\bigtriangleup v}{\left(P^+\right)^{3/2}}\sqrt{\frac{x_3}{2x_1x_2}}\Vec{\epsilon}^\star_{\perp,1}\cdot \left[\left(\frac{\Vec{q}_{\perp}}{x}\right)_1-\left(\frac{\Vec{q}_{\perp}}{x}\right)_3\right]\delta^{\lambda_3}_{-\lambda_2} \enskip \enskip i f^{a_1}_{a_2a_3}$ \\   
        \hline
        $V_{4.2}$ &  $-\frac{\bigtriangleup v}{\left(P^+\right)^{3/2}}\sqrt{\frac{x_2}{2x_1x_3}}\Vec{\epsilon}_{\perp,3}\cdot \left[\left(\frac{\Vec{q}_{\perp}}{x}\right)_3-\left(\frac{\Vec{q}_{\perp}}{x}\right)_2\right]\delta^{\lambda_2}_{\lambda_1} \enskip \enskip i f^{a_1}_{a_2a_3}$ \\   
        \hline
        $V_{4.3}$ &  $-\frac{\bigtriangleup v}{\left(P^+\right)^{3/2}}\sqrt{\frac{x_1}{2x_2x_3}}\Vec{\epsilon}_{\perp,3}\cdot \left[\left(\frac{\Vec{q}_{\perp}}{x}\right)_3-\left(\frac{\Vec{q}_{\perp}}{x}\right)_1\right]\delta^{\lambda_2}_{\lambda_1} \enskip \enskip i f^{a_1}_{a_2a_3}$ \\   
        \hline
    \end{tabular}

\end{table}

The factor $\frac{g P^+}{\sqrt{2(2\pi)^3}}$ included in $\Delta v$ in table~\ref{tab:V}
changes upon discretizing. The same applies to all the factors $2(2\pi)^3$ in the rest of this
document, which change, upon taking momentum to a finite grid, in the same manner, to $2^3\times 2^{N_\text{longitudinal}}\times \pi$, with $N_\text{longitudinal}$ the number of qubits assigned to longitudinal momentum.

In this work, we set the total longitudinal momentum to $P^+=10\, \mathrm{GeV}$, at which scale we take $\alpha_s(P^+)=0.1791$. 

The full set of parameters employed is summarized in tables~\ref{tab:parameters_General} and ~\ref{tab:parameters_specific}.

\begin{table}
    \centering
        \caption{Generic parameters controlling the calculation. $P^+$ sets the scale of the simulation; $\alpha_s$ the coupling constant $g = \sqrt{4\pi \alpha_s}$;  $m_c$ is the charm-quark mass;  longitudinal momentum fractions (m.f.) $x$ or $z$ from $p^+ = x P^+$ fall on an equispaced grid;
        and $\kappa$  is related to the width of the $J/\Psi$ wavefunction. All energy dimensions are in GeV.}
    \label{tab:parameters_General}
    \begin{tabular}{|c|c|c|c|c|c|c|c|c|c|}
    \hline
    $P^+$ & $\alpha_s(P^+)$ & $m_c$ & $x_i$ &  $\kappa$  \\
    \hline\hline
     $10$ &  $0.18$ & $1.27$ & $k/(N+1), k=1,\dots,N$ & $1.34$ \\
    \hline    

    \end{tabular}
\end{table}

\begin{table}
    \centering
    \caption{Specific parameters of each simulation. They include either kinetic and vertex interactions (V) or the totality of possible terms within the Fock space truncation (T). $N_c$ is the number of colors, $N$ the number of available longitudinal momentum fractions $z_i$ in the grid, $N_ {qubits}$ the total number of qubits; $\Delta t$ is the Trotter step; $\epsilon$ is a size cutoff on the potential matrix elements, see main text or section \ref{sec:scaling} below; and the last column is the execution time per shot and Trotter step on the PROTEUS \textit{Nix} supercomputer~({\tt  https://proteus.ugr.es/}), employing 4 nodes at 48 cores per node and EPYC 9274F architecture.}
\label{tab:parameters_specific}
    \begin{tabular}{|c|c|c|c|c|c|c|c|}
        \hline
         Run           & $V/T$ & $N_c$ & $N$ & $N_{\text{qubits}}$ & $\Delta t $ & $\epsilon$ & Runtime \\ 
         configuration & & & grid & &(GeV$^{-1}$) & & (min)
         \\ 
        \hline\hline
        1 & V & 2 & 4 & 25 & 0.2 & 0 & 1.5\\ 
        \hline
        2 & T & 2 & 4 & 25 & 0.1 & 0 & 22 \\
        \hline
        3 & T & 2 & 4 & 25 & 0.2 & 0.02 & 15 \\ 
        \hline
        4 & V & 2 & 8 & 29 & 0.2 & 0 & 110 \\ 
        \hline
        5 & T & 2 & 8 & 29 & 0.2 & 0.02 & 1800 \\ 
        \hline
        6 & V & 3 & 4 & 29 & 0.25 & 0 & 120 \\ 
        \hline

    \end{tabular}
\end{table}

%%%%%%%%%%%%%%%%%%%%%%%%%%%%%%%%%%%%%%%%%%%%%%
\subsection{Other interactions}
%%%%%%%%%%%%%%%%%%%%%%%%%%%%%%%%%%%%%%%%%%%%%%

Next, we read off the seagull, $S$, contracted terms $C$ and fork--interactions $F$.

\begin{table}
    \centering
        \caption{The ``Seagull'' interaction in which four particles touch at the same spacetime point, in terms of the Dirac spinors and polarization vectors, with  $\bigtriangleup = \hat{g}^2 \delta\left(q^+_1+q^+_2-q^+_3 -q^+_4 \right) \delta^{(2)}\left(\Vec{q}_{\perp 1}+\Vec{q}_{\perp 2}-\Vec{q}_{\perp 3}-\Vec{q}_{\perp 4} \right)$, and with  $\hat{g}^2= \frac{g^2 P^+ }{4\left(2 \pi \right)^3}.$}
    \label{tab:S}
    \begin{tabular}{|c|c|}
     \hline
     $S_1$ & $-\frac{\bigtriangleup }{\sqrt{q^+_1q^+_2q^+_3q^+_4}} \frac{\left( \overline{u}_1T^a\gamma^+ u_3  \right)\left( \overline{u}_2\gamma^+T^a u_4  \right)}{\left( q^+_1-q^+_3\right)^2}$ \\     \hline
     $S_{3,1}$ & $+\frac{2\bigtriangleup }{\sqrt{q^+_1q^+_2q^+_3q^+_4}} \frac{\left( \overline{u}_1T^a\gamma^+ u_3  \right)\left( \overline{v}_4\gamma^+T^a v_2  \right)}{\left( q^+_1-q^+_3\right)^2}$ \\     \hline
     $S_{3,2}$ & $-\frac{2\bigtriangleup }{\sqrt{q^+_1q^+_2q^+_3q^+_4}} \frac{\left( \overline{u}_1T^a\gamma^+ v_2  \right)\left( \overline{v}_4\gamma^+T^a u_3  \right)}{\left( q^+_1+q^+_2\right)^2}$ \\     \hline
     $S_{4,1}$&  $+\frac{\bigtriangleup }{\sqrt{q^+_1q^+_2q^+_3q^+_4}} \frac{\left( \overline{u}_1 T^{a_4} \cancel{\epsilon}_4\gamma^+\cancel{\epsilon}^\star_2 T^{a_2} u_3  \right)}{\left( q^+_1-q^+_4\right)}$   \\  \hline
     $S_{4,2}$&  $+\frac{\bigtriangleup }{\sqrt{q^+_1q^+_2q^+_3q^+_4}} \frac{\left( \overline{u}_1 T^{a_2} \cancel{\epsilon}^\star_2\gamma^+\cancel{\epsilon}_4 T^{a_4} u_3  \right)}{\left( q^+_1+q^+_2\right)}$   \\  \hline       
     $S_{4,3}$&    $+\frac{2 \left(q^+_2+q^+_4\right)\bigtriangleup }{\sqrt{q^+_1q^+_2q^+_3q^+_4}} \frac{\left( \overline{u}_1T^a\gamma^+ u_3  \right)\left( \epsilon^\star_2 i f^a \epsilon_4  \right)}{\left( q^+_1-q^+_3\right)^2}$  \\     \hline
     $S_{6,1}$&  $+\frac{\bigtriangleup }{\sqrt{q^+_1q^+_2q^+_3q^+_4}} \frac{\left( \overline{u}_1 T^{a_3} \cancel{\epsilon}_3\gamma^+\cancel{\epsilon}_4 T^{a_4} v_2  \right)}{\left( q^+_1-q^+_3\right)}$   \\  \hline
     $S_{6,2}$&    $-\frac{ \left(q^+_3-q^+_4\right)\bigtriangleup }{\sqrt{q^+_1q^+_2q^+_3q^+_4}} \frac{\left( \overline{u}_1T^a\gamma^+ v_2  \right)\left( \epsilon_3 i f^a \epsilon_4  \right)}{\left( q^+_1+q^+_2\right)^2}$  \\     \hline
     $S_{7,1}$&    $-\frac{ \left(q^+_1+q^+_3\right)\left(q^+_2+q^+_4\right)\bigtriangleup }{\sqrt{q^+_1q^+_2q^+_3q^+_4}} \frac{\left( \epsilon^\star_1  f^a \epsilon_3   \right)\left( \epsilon^\star_2  f^a \epsilon_4  \right)}{\left( q^+_1-q^+_3\right)^2}$  \\     \hline
     $S_{7,2}$&    $+\frac{2 q^+_3q^+_4\bigtriangleup }{\sqrt{q^+_1q^+_2q^+_3q^+_4}} \frac{\left( \epsilon^\star_1  f^a \epsilon^\star_2   \right)\left( \epsilon_3  f^a \epsilon_4  \right)}{\left( q^+_1+q^+_2\right)^2}$  \\     \hline
     $S_{7,3}$&    $+\frac{ \bigtriangleup }{\sqrt{q^+_1q^+_2q^+_3q^+_4}} \left( \epsilon^\star_1   \epsilon_3   \right)\left( \epsilon^\star_2   \epsilon_4  \right) f^a_{a_1a_2} f^a_{a_3a_4}$  \\     \hline
     $S_{7,4}$&    $+\frac{ \bigtriangleup }{\sqrt{q^+_1q^+_2q^+_3q^+_4}} \left( \epsilon^\star_1   \epsilon_3   \right)\left( \epsilon^\star_2   \epsilon_4  \right) f^a_{a_1a_4} f^a_{a_3a_2}$  \\     \hline
     $S_{7,5}$&    $+\frac{ \bigtriangleup }{\sqrt{q^+_1q^+_2q^+_3q^+_4}} \left( \epsilon^\star_1   \epsilon^\star_2   \right)\left( \epsilon_3   \epsilon_4  \right) f^a_{a_1a_3} f^a_{a_2a_4}$  \\     \hline
    \end{tabular}
\end{table}

\begin{table}
    \centering
        \caption{The Fork interactions $1\rightarrow3$ written down in terms of Dirac spinors. Here, $\bigtriangleup =\frac{g^2 P^+ }{4\left(2 \pi \right)^3}  \delta\left(q^+_1-q^+_2-q^+_3 -q^+_4 \right) \delta^{(2)}\left(\Vec{q}_{\perp 1}-\Vec{q}_{\perp 2}-\Vec{q}_{\perp 3}-\Vec{q}_{\perp 4} \right)$. }
    \label{tab:F}
    \renewcommand{\arraystretch}{1.7}
    \begin{tabular}{|c|c|}
    \hline
        $F_1$ & $+\frac{2\bigtriangleup \left( \overline{u}_1T^a\gamma^+ u_2  \right)\left( \overline{v}_3\gamma^+T^a u_4  \right)}{\sqrt{q^+_1q^+_2q^+_3q^+_4} \left( q^+_1-q^+_2\right)^2}$ \\ \hline

        $F_{3,1}$ & $+\frac{ \bigtriangleup \left( \overline{u}_1 T^{a_4} \cancel{\epsilon}_4\gamma^+\cancel{\epsilon}_3 T^{a_3} u_2  \right)}{\sqrt{q^+_1q^+_2q^+_3q^+_4} \left( q^+_1-q^+_4\right)}$ \\ \hline

        $F_{3,2}$ & $-\frac{2 q^+_3 \bigtriangleup \left( \overline{u}_1T^a\gamma^+ u_2  \right)\left( \epsilon_3 i f^a \epsilon_4  \right)}{\sqrt{q^+_1q^+_2q^+_3q^+_4} \left( q^+_1-q^+_2\right)^2}$ \\ \hline

        $F_{5,1}$ & $+\frac{ \bigtriangleup \left( \overline{v}_3 T^{a_1} \cancel{\epsilon}_1^\star \gamma^+\cancel{\epsilon}_2 T^{a_2} u_4  \right)}{\sqrt{q^+_1q^+_2q^+_3q^+_4} \left( q^+_1-q^+_3\right)}$ \\ \hline

        $F_{5,2}$ & $-\frac{ \bigtriangleup \left( \overline{v}_3 T^{a_2} \cancel{\epsilon}_2 \gamma^+\cancel{\epsilon}_1^\star T^{a_1} u_4  \right)}{\sqrt{q^+_1q^+_2q^+_3q^+_4} \left( q^+_1-q^+_4\right)}$ \\ \hline

        $F_{5,3}$ & $+\frac{2 \left(q^+_1+q^+_2\right)\bigtriangleup \left( \overline{v}_3T^a\gamma^+ u_4  \right)\left( \epsilon^\star_1 i f^a \epsilon_2  \right)}{\sqrt{q^+_1q^+_2q^+_3q^+_4} \left( q^+_1-q^+_2\right)^2}$ \\ \hline

        $F_{6,1}$ & $+\frac{2 q^+_3\left(q^+_1+q^+_2\right)\bigtriangleup \left( \epsilon^\star_1  f^a \epsilon_2   \right)\left( \epsilon_3  f^a \epsilon_4  \right)}{\sqrt{q^+_1q^+_2q^+_3q^+_4} \left( q^+_1-q^+_2\right)^2}$ \\ \hline

        $F_{6,2}$ & $+\frac{2 \bigtriangleup \left( \epsilon^\star_1   \epsilon_3   \right)\left( \epsilon_2   \epsilon_4  \right) f^a_{a_1a_2} f^a_{a_3a_4}}{\sqrt{q^+_1q^+_2q^+_3q^+_4}}$ \\ \hline

    \end{tabular}
\end{table}

\begin{table}
    \centering
        \caption{$C_{n,j}$ coefficients of the contracted operators yielding contributions to the self--energy in terms of matrix elements of the Seagull $S$ terms.}
    \label{tab:C}
    \resizebox{8cm}{!} {
    \begin{tabular}{|c|c|}
    \hline
        $C_{1,1}\left(q_1\right)$ &  $\sum_{2}\left( S_{1}\left(1,2;2,1\right)+S_{3,2}\left(1,2;1,2\right)\right)$ \\     \hline
        $C_{1,2}\left(q_1\right)$ &  $\sum_{2}\left( S_{4,1}\left(1,2;2,1\right)+S_{4,2}\left(1,2;1,2\right)\right) \left(\frac{1}{2}\right)$ \\     \hline
        $C_{3,1}\left(q_1\right)$ &  $\sum_{2}\left( S_{4,1}\left(2,1;2,1\right)+S_{4,2}\left(2,1;2,1\right)\right)\left(-\frac{1}{2}\right)$\\     \hline
        $C_{3,2}\left(q_1\right)$ &  $\sum_{2}\left( S_{7,1}\left(1,2;2,1\right)+S_{7,2}\left(1,2;2,1\right)\right)$\\     \hline
        $C_{3,3}\left(q_1\right)$ &  $\sum_{2}\left( S_{7,3}\left(1,2;1,2\right)+S_{7,3}\left(1,2;2,1\right) 
        
        +S_{7,4}\left(1,2;1,2\right)+S_{7,5}\left(1,2;2,1\right)\right)$\\     \hline
    \end{tabular}
    }
\end{table}

Tables  \ref{tab:F}, \ref{tab:S} and \ref{tab:C} detail all  $F$, $S$ and $C$ terms of colors and helicities.
Moreover, in tables~\ref{tab:F_x}, \ref{tab:S_x} and \ref{tab:C_x}, we display $F$, $S$ and $I$ in terms of helicities, color, intrinsic momentum and flavor, with  $I$ a shorthand for $C$ absorbing a factor of longitudinal momentum fraction $x$ as follows,
\begin{equation}
I_{i,q} = x_{iq} C_i\left(q\right)  
\end{equation}

\begin{table}
    \centering
        \caption{Fork interactions in terms of helicities, color and flavor
    Here, $\bigtriangleup = \frac{g^2 P^+ }{4\left(2 \pi \right)^3} \delta\left(q^+_1-q^+_2-q^+_3 -q^+_4 \right) \delta^{(2)}\left(\Vec{q}_{\perp ,1}-\Vec{q}_{\perp ,2}-\Vec{q}_{\perp ,3}-\Vec{q}_{\perp ,4} \right)$. }
    \label{tab:F_x}
    \begin{tabular}{|c|c|}
    \hline
        $F_1$ & $+\frac{2\bigtriangleup }{\left(P^+\right)^2\left( x_1-x_2\right)^2}\enskip\delta^{\lambda_2}_{\lambda_1}\delta^{\lambda_4}_{-\lambda_3} \enskip\delta^{l_2}_{l_1}\delta^{l_4}_{l_3}\enskip  T^a_{c_1c_2}T^a_{c_3c_4}$ \\     \hline
         $F_{3,1}$&  $+\frac{\bigtriangleup }{\left(P^+\right)^2\left( x_1-x_4\right)\sqrt{x_3x_4}}\enskip\delta^{\lambda_2}_{\lambda_1}\delta^{\lambda_4}_{-\lambda_3}\delta^{\lambda_4}_{\lambda_1} \enskip \delta^{l_2}_{l_1}\enskip T^{a_4}_{c_1c}T^{a_3}_{cc_2}$ \\  \hline
         $F_{3,2}$&     $+\frac{\bigtriangleup }{\left(P^+\right)^2\left( x_1-x_2\right)^2}\sqrt{\frac{x_3} {x_4}}\enskip \delta^{\lambda_2}_{\lambda_1}\delta^{\lambda_4}_{-\lambda_3} \enskip\delta^{l_2}_{l_1} \enskip T^{a}_{c_1c_2}f^{a}_{a_3a_4}$ \\     \hline
          $F_{5,1}$& $+\frac{\bigtriangleup }{\left(P^+\right)^2\left( x_1-x_3\right)\sqrt{x_1x_2}}\enskip\delta^{\lambda_2}_{\lambda_1}\delta^{\lambda_4}_{-\lambda_3}\delta^{\lambda_4}_{-\lambda_1} \enskip \delta^{l_4}_{l_3}\enskip T^{a_1}_{c_3c}T^{a_2}_{cc_4}$   \\  \hline
          $F_{5,2}$& $-\frac{\bigtriangleup }{\left(P^+\right)^2\left( x_1-x_4\right)\sqrt{x_1x_2}}\enskip\delta^{\lambda_2}_{\lambda_1}\delta^{\lambda_4}_{-\lambda_3}\delta^{\lambda_4}_{\lambda_1}  \enskip\delta^{l_4}_{l_3} \enskip T^{a_2}_{c_3c}T^{a_1}_{cc_4}$     \\  \hline
          $F_{5,3}$&     $+\frac{\bigtriangleup \left(x_1+x_2\right)}{\left(P^+\right)^2\left( x_1-x_2\right)^2\sqrt{x_1x_2}}\enskip\delta^{\lambda_2}_{\lambda_1}\delta^{\lambda_4}_{-\lambda_3}\enskip \delta^{l_4}_{l_3}\enskip i f^{a}_{a_1a_2}T^{a}_{c_3c_4}$  \\     \hline
          $F_{6,1}$&    $+\frac{\bigtriangleup \left(x_1+x_2\right)x_3}{2\left(P^+\right)^2\left( x_1-x_2\right)^2\sqrt{x_1x_2x_3x_4}}\enskip\delta^{\lambda_2}_{\lambda_1}\delta^{\lambda_4}_{-\lambda_3}\enskip \enskip i f^{a}_{a_1a_2}f^{a}_{c_3c_4}$   \\     \hline
          $F_{6,2}$&  $+\frac{\bigtriangleup }{2\left(P^+\right)^2\sqrt{x_1x_2x_3x_4}}\enskip\delta^{\lambda_3}_{\lambda_1}\delta^{\lambda_4}_{-\lambda_2}\enskip \enskip i f^{a}_{a_1a_2}f^{a}_{c_3c_4}$ \\     \hline
 \end{tabular}
\end{table}
\begin{table}
    \centering
        \caption{The seagull interaction in terms of helicities, color and flavor. In this case,  $\bigtriangleup = \frac{g^2 P^+ }{4\left(2 \pi \right)^3} \delta\left(q^+_1+q^+_2-q^+_3 -q^+_4 \right) \delta^{(2)}\left(\Vec{q}_{\perp ,1}+\Vec{q}_{\perp ,2}-\Vec{q}_{\perp ,3}-\Vec{q}_{\perp ,4} \right)$.}
    \label{tab:S_x}
    \begin{tabular}{|c|c|}
     \hline
     $S_1$ & $+\frac{-\bigtriangleup }{\left(P^+\right)^2\left( x_1-x_3\right)^2}\enskip\delta^{\lambda_3}_{\lambda_1}\delta^{\lambda_4}_{\lambda_2} \enskip\delta^{l_3}_{l_1}\delta^{l_4}_{l_2}\enskip  T^a_{c_1c_3}T^a_{c_2c_4}$ \\     \hline
     $S_{3,1}$ & $+\frac{2\bigtriangleup }{\left(P^+\right)^2\left( x_1-x_3\right)^2}\enskip\delta^{\lambda_3}_{\lambda_1}\delta^{\lambda_4}_{\lambda_2} \enskip\delta^{l_3}_{l_1}\delta^{l_4}_{l_2}\enskip  T^a_{c_1c_3}T^a_{c_4c_2}$ \\     \hline
     $S_{3,2}$ & $-\frac{2\bigtriangleup }{\left(P^+\right)^2\left( x_1+x_2\right)^2}\enskip\delta^{\lambda_2}_{-\lambda_1}\delta^{\lambda_4}_{-\lambda_3} \enskip\delta^{l_2}_{l_1}\delta^{l_4}_{l_3}\enskip  T^a_{c_1c_2}T^a_{c_4c_3}$ \\     \hline
     $S_{4,1}$&  $+\frac{\bigtriangleup }{\left(P^+\right)^2\left( x_1-x_4\right)\sqrt{x_2x_4}}\enskip\delta^{\lambda_3}_{\lambda_1}\delta^{\lambda_4}_{\lambda_2}\delta^{\lambda_2}_{\lambda_1} \enskip \delta^{l_3}_{l_1}\enskip T^{a_4}_{c_1c}T^{a_2}_{cc_3}$   \\  \hline
     $S_{4,2}$&  $+\frac{\bigtriangleup }{\left(P^+\right)^2\left( x_1+x_2\right)\sqrt{x_2x_4}}\enskip\delta^{\lambda_3}_{\lambda_1}\delta^{\lambda_4}_{\lambda_2}\delta^{\lambda_2}_{-\lambda_1} \enskip \delta^{l_3}_{l_1}\enskip T^{a_2}_{c_1c}T^{a_4}_{cc_3}$   \\  \hline       
     $S_{4,3}$&     $+\frac{\bigtriangleup \left(x_2+x_4\right)}{\left(P^+\right)^2\left( x_1-x_3\right)^2\sqrt{x_2x_4}}\enskip\delta^{\lambda_3}_{\lambda_1}\delta^{\lambda_4}_{\lambda_2}\enskip \delta^{l_3}_{l_1}\enskip i T^{a}_{c_1c_3}f^{a}_{a_2a_4}$ \\     \hline
     $S_{6,1}$& $+\frac{\bigtriangleup }{\left(P^+\right)^2\left( x_1-x_3\right)\sqrt{x_3x_4}}\enskip\delta^{\lambda_2}_{-\lambda_1}\delta^{\lambda_4}_{-\lambda_3}\delta^{\lambda_3}_{\lambda_1} \enskip \delta^{l_2}_{l_1}\enskip T^{a_3}_{c_1c}T^{a_4}_{cc_2}$     \\  \hline
     $S_{6,2}$&   $+\frac{\bigtriangleup \left(x_3-x_4\right)}{\left(P^+\right)^2\left( x_1+x_2\right)^2\sqrt{x_3x_4}}\enskip\delta^{\lambda_2}_{-\lambda_1}\delta^{\lambda_4}_{-\lambda_3}\enskip \delta^{l_2}_{l_1}\enskip i T^{a}_{c_1c_2}f^{a}_{a_3a_4}$  \\     \hline
     $S_{7,1}$&   $-\frac{\bigtriangleup \left(x_1+x_3\right)\left(x_2+x_4\right)}{4\left(P^+\right)^2\left( x_1-x_3\right)^2\sqrt{x_1x_2x_3x_4}}\enskip\delta^{\lambda_3}_{\lambda_1}\delta^{\lambda_4}_{\lambda_2}\enskip \enskip  f^{a}_{a_1a_3}f^{a}_{a_2a_4}$  \\     \hline
     $S_{7,2}$&     $\frac{\bigtriangleup }{2\left(P^+\right)^2\left( x_1+x_2\right)^2}\sqrt{\frac{x_1x_3}{x_2x_4}}\enskip\delta^{\lambda_2}_{-\lambda_1}\delta^{\lambda_4}_{-\lambda_3}\enskip \enskip  f^{a}_{a_1a_2}f^{a}_{a_3a_4}$   \\     \hline
     $S_{7,3}$&     $\frac{\bigtriangleup }{4\left(P^+\right)^2\sqrt{x_1x_2x_3x_4}}\enskip\delta^{\lambda_3}_{\lambda_1}\delta^{\lambda_4}_{\lambda_2}\enskip \enskip  f^{a}_{a_1a_2}f^{a}_{a_3a_4}$  \\     \hline
     $S_{7,4}$&     $\frac{\bigtriangleup }{4\left(P^+\right)^2\sqrt{x_1x_2x_3x_4}}\enskip\delta^{\lambda_2}_{-\lambda_1}\delta^{\lambda_4}_{-\lambda_3}\enskip \enskip  f^{a}_{a_1a_3}f^{a}_{a_2a_4}$ \\     \hline
     $S_{7,5}$&     $\frac{\bigtriangleup }{4\left(P^+\right)^2\sqrt{x_1x_2x_3x_4}}\enskip\delta^{\lambda_3}_{\lambda_1}\delta^{\lambda_4}_{\lambda_2}\enskip \enskip  f^{a}_{a_1a_4}f^{a}_{a_3a_2}$ \\     \hline
    \end{tabular}
\end{table}

\begin{table}[h!]
    \centering
        \caption{$I_{n,j}$ coefficients of the contracted operators $C_i$ from the corresponding seagull matrix elements. Here $\bar{g} = g^2/(2\pi)^3$.}
    \label{tab:C_x}
    \resizebox{8cm}{!} {
    \begin{tabular}{|c|c|}
    \hline
        $I_{1,1}\left(q_1\right)$ & $\overline{g}^2\frac{\left(N^2_c-1\right)}{2N_c}\sum_{x,\Vec{q}_\perp}\left[\frac{x_1}{\left(x_1-x\right)^2}-\frac{x_1}{\left(x_1+x\right) ^2} \right ]$ \\     \hline
        $I_{1,2}\left(q_1\right)$ &  $\overline{g}^2\frac{\left(N^2_c-1\right)}{4N_c}\sum_{x,\Vec{q}_\perp}\left[\frac{x_1}{x_1-x}+\frac{x_1}{x_1+x} \right ]\frac{1}{x}$\\     \hline
        $I_{3,1}\left(q_1\right)$ & $\overline{g}^2\frac{N_f}{2}\sum_{x,\Vec{q}_\perp}\left[\frac{1}{x_1-x}-\frac{1}{x_1+x} \right ]$\\     \hline
        $I_{3,2}\left(q_1\right)$ &  $\overline{g}^2\frac{N_c}{4}\sum_{x,\Vec{k_\perp}}\left[\frac{\left(x_1+x\right)^2}{\left(x_1-x\right)^2}+\frac{\left(x_1-x\right)^2}{\left(x_1+x\right)^2} \right ]\frac{1}{x}$ \\     \hline
        $I_{3,3}\left(q_1\right)$ &  $\overline{g}^2\frac{N_c}{2}\sum_{x,\Vec{q}_\perp}\frac{1}{x}$ \\     \hline
    \end{tabular}
    }
\end{table}

%%%%%%%%%%%%%%%%%%%%%%%%%%%%%%%%%%%%%%%%%%%%%%%%%%%%%%%%%%%%%%
\subsection{Minor differences with published material}
%%%%%%%%%%%%%%%%%%%%%%%%%%%%%%%%%%%%%%%%%%%%%%%%%%%%%%%%%%%%%%
We here point out four very minor differences respect to the standard reference in the field, published by S. J. Brodsky, H.-C. Pauli and S. S. Pinsky in Phys.Rept. {\bf 301} (1998), pages 299-486 (shortened to BP\&P). None of them is major and none of them affects the computation that we here report, because to start with the $c$ quark leading a jet, to fill a memory which can fit at most two quarks, one antiquark and one gluon, none of these following terms can actually contribute. But for future reference and completeness, we list the differences between our tables and theirs.

First, the term $F_{62}$, corresponding to the non-Abelian four-gluon vertex in the time order of a trident, $g_1\to g_2g_3g_4$ reads in Brodsky, Pauli and Pinsky (BP\&P) as follows:
\begin{equation}
F_{62} = \frac{\Delta}{2\sqrt{x_1x_2x_3x_4}} \delta_{\lambda_1}^{\lambda_3} \delta_{\lambda_2}^{\lambda_4} C^a_{a_1a_2} C^a_{a_3a_4}
\end{equation}
(in more standard notation, $C^a_{bc}$, the structure constant of $SU(3)$, is typed as $f^{abc}$). The sign of the gluon polarization
should, in our view, be changed, $\lambda_2\to - \lambda_2$, that is, $\delta_{\lambda_2}^{\lambda_4}\to \delta_{-\lambda_2}^{\lambda_4}$. This seems consistent with the rest of the same vertex, is probably a typo and deserves no further comment.

Second, we report a small difference which also seems to be a typo in the term $F_{31}$ which represents a fermion line $1\to 2$ emitting two gluons at two different points which are  ``simultaneous'' in the light-front-time sense. In BP\&P this reads
\begin{equation}
F_{31}= \frac{\Delta}{(x_1-x_4)}\frac{1}{\sqrt{x_3x_4}} \delta_{\lambda_1}^{\lambda_2} \delta_{-\lambda_3}^{\lambda_4}
\delta_{\lambda_1}^{\lambda_4}\delta_{f_1}^{f_2} T^{a_3}_{c_1c} T^{a_4}_{cc_2}\ .  
\end{equation}
We believe that the superindices $a_3$ and $a_4$ of the color matrices should be exchanged to read 
$T^{a_4}_{c_1c} T^{a_3}_{cc_2}$ instead, given the order of the emissions, $(1\to 4+\text{intermediate\ c})$ followed by 
$(\text{intermediate\ c}\to 2+3)$.

In the third place, we observe that the square root of products of particle momentum fractions, $\sqrt{\prod x_i}$ perhaps contains a typo in BP\&P too, due to the exchange of pieces between the terms $V_{42}$ and $V_{43}$. 

Finally, we multiplied by relative  $1/P^+$ factors absent in BP\&P in passing between tables \ref{tab:V} and \ref{tab:V_x}; \ref{tab:F} and \ref{tab:F_x} and \ref{tab:S} and \ref{tab:S_x}.

%%%%%%%%%%%%%%%%%%%%%%%%%%%%%%%%%%%%%%%%%%%%%%%%%%%%%%%%%%%%%%%%%%%%
\section{Detailing the encoding}\label{app:encoding}
%%%%%%%%%%%%%%%%%%%%%%%%%%%%%%%%%%%%%%%%%%%%%%%%%%%%%%%%%%%%%%%%%%%%
\subsection{ Organization of the (quantumlike) memory}
%%%%%%%%%%%%%%%%%%%%%%%%%%%%%%%%%%%%%%%%%%%%%%
The quantum memory is organized in registers sized to the particle type, whether quark/antiquark or gluon. One qubit within each register is assigned to track the presence or absence of a particle therein, and the others the eventual--particle quantum numbers. The presence qubit is, in our common practice, used as a control, changing from $\ket{1}$ (presence) to $\ket{0}$ (absence) upon creation or annihilation; the remaining qubits are transformed by purpose--coded \textit{set}, $\set{\alpha} = \ket{\alpha}\bra{0...0}$ and \textit{scrap}, $\scrap{\alpha}=\ket{0...0}\bra{\alpha}$ operators. The mapping between the Hamiltonian creation and annihilation operators and the basic quantum gates is realized by 
\begin{align}
a^{(n)\dagger }_{\alpha} & =  \sum^{n}_{j = 1}  \mathcal{S}_{j} \cdot \projector{n-j}{0}\otimes \left(\ketbrac{1}{0}\otimes \mathfrak{s}^{\dagger}_{\alpha}\right)_{j} \otimes \,\projector{j-1}{j-1}\\
b^{(n)\dagger }_{\alpha} &=  \sum^{n}_{j = 1}\mathcal{A}_{j}\cdot \projector{n-j}{0}\otimes \left(\ketbrac{1}{0}\otimes \mathfrak{s}^{\dagger}_{\alpha}\right)_{j}
\otimes \,\projector{j-1}{j-1},
    \label{def:nReg-bosoncreation}
\end{align}
for bosonic and fermionic degrees of freedom, respectively. The projectors $\projector{n}{i}$ act on $n$ registers and unless exactly $i$ registers are occupied, they project to zero. The control operators $\ketbrac{i}{j}$ transform the presence qubits from $\ket{j}$ to $\ket{i}$ (absence $\leftrightarrow$ presence). The step symmetrizers $\mathcal{S}_{j}$ symmetrize the $j$ register with the remaining $j-1$ registers of the same corresponding particle species; the operators $\mathcal{A}_{j}$ antisymmetrize the memory in a similar way. Thus, in our encoding, the basis states are particle configurations instead of modes as in other types of implementation. 

Light-front time evolution requires the exponentiation of the Hamiltonian terms of the preceding section, written as sums of vertex functions and product of creation and annihilation operators. As an example consider the exponential of the fermion-gluon scattering vertex
\begin{equation}
    U_1(\Delta t) = \exp\left\{-i\Delta t\sum_{1,2,3} V_1\left(1;2,3\right) b^{\dagger}_1b_2 a_3 + h.c.\right\},
    \label{unitaryop}
\end{equation}
which conserves the number of fermions but not of gluons. In our implementation the maximum number of fermions and gluons is $n_f =2$ and $n_g = 1$, so that
\begin{align}
b^{\dagger}_1&b_2 a_3 = \projector{1_f}{0}\otimes \left(\ketbrac{1}{1}\otimes\set{1}\scrap{2}\right)_{1_f}\otimes     \left(\ketbrac{0}{1}\otimes\scrap{3}\right)_{1_g}\nonumber\\
+&\mathcal{A}_{2_f}\cdot\left(\ketbrac{1}{1}\otimes\set{1}\scrap{2}\right)_{2_f}\otimes\projector{1_f}{1}\otimes\left(\ketbrac{0}{1}\otimes\scrap{3}\right)_{1_g}\cdot \mathcal{A}_{2_f},
\label{V1_ss}
\end{align}
the equation can be further simplified if the memory is constrained to be filled in order, but we will not delve into the details here. When exponentiating, the term in the first row acts when there is only one fermion in the memory, while the second row acts only when there are two. Thus, the antisymmetrization operators are only necessary on the second row.  
Details about exponentiation of equations such as Eq.~(\ref{V1_ss}) have already been provided in a separate, methodological publication~\cite{Galvez-Viruet:2024hry}. 

In brief, we first define and code a unitary version of the antisymmetrizer operator $\hat{\mathcal{A}}^{\dagger}_2(\alpha,\neg\beta)$ that antisymmetrizes the second register only if it stores mode $\alpha$ and the first register does not store mode $\beta$:
\begin{align}
    \hat{\mathcal{A}}^{\dagger}_2(\alpha,\neg \beta)\ket{\gamma}\ket{\delta} =\left\{
	\begin{array}{ll}
		\frac{1}{\sqrt{2}}\left(\ket{\alpha}\ket{\delta}-\ket{\delta}\ket{\alpha}\right)  & \mbox{if } \alpha = \gamma,\beta\,\neq\delta \\
		\ket{\gamma}\ket{\delta} & \mbox{otherwise. }
	\end{array}
\right.
\end{align}
The inverse procedure defines $\hat{\mathcal{A}}_2(\alpha,\neg \beta)$. 

Then, using the Trotter decomposition to separately exponentiate both rows of Eq.(\ref{V1_ss}) and to transform the sum over modes into a product one arrives at
\begin{equation}
    U_1(\Delta t) = \projector{1_f}{0}\otimes U_{1,n_f = 1}(\Delta t)\,\,U_{1,n_f = 2}(\Delta t)\otimes\projector{1_f}{1} ,
\end{equation}
with 
\begin{equation}
    U_{1,1}(\Delta t) =\prod_{123}e^{ -i\Delta t \,V_1\left\{\left(\ketbrac{1}{1}\otimes\set{1}\scrap{2}\right)_{1_f}\otimes     \left(\ketbrac{0}{1}\otimes\scrap{3}\right)_{1_g}+h.c.\right\}},
     \label{eq:U11term}
\end{equation}
and 
\begin{align}
    U_{1,2}(\Delta t) =&\prod_{123}\hat{\mathcal{A}}^{\dagger}_2(1)\hat{\mathcal{A}}^{\dagger}_2(2,\neg 1)\nonumber\\
    &e^{ -i\Delta t \,V_1\left\{\left(\ketbrac{1}{1}\otimes\set{1}\scrap{2}\right)_{2_f}\otimes     \left(\ketbrac{0}{1}\otimes\scrap{3}\right)_{1_g}+h.c.\right\}_{\neg 12}}\nonumber\\
    &\hat{\mathcal{A}}_2(2,\neg 1)\hat{\mathcal{A}}_2(1),
    \label{eq:U12term}
\end{align}
the subscript $\neg 12$ indicates that the rotation must be avoided if modes $1$ and $2$ are both stored in memory, thus preventing repetition, forbidden by fermion statistics. The operator $\hat{\mathcal{A}}_2(1)$ extract mode $1$ from an antisymmetric state, and $\hat{\mathcal{A}}_2(2,\neg 1)$ extracts mode 
$2$ provided mode $1$ has not been already extracted. After the extractions the rotations are applied using Gray codes and multi-controlled gates. Finally, the result is antisymetrized again. Similar decompositions and operators are used for the exponentiation of the remaining Hamiltonian terms.

%%%%%%%%%%%%%%%%%%%%%%%%%%%%%%%%%%%%%%%%%%%%%%%%%%%%%%%%%%%%%%%%%%%%
\subsection{ Momentum grid  }
%%%%%%%%%%%%%%%%%%%%%%%%%%%%%%%%%%%%%%%%%%%%%%%%%%%%%%%%%%%%%%%%%%%
\label{subsection:momentumgrid}
In principle, for each of the three dimensions on the quantization hypersurface, we would have $i=1\dots N$ momentum values, with $N=2^n$ the number of modes. We have taken $n=0$ for the transverse ${\bf k}_\perp$ directions and $n=3$, so that $N=8$, in the longitudinal direction (but these are variables in the computer code, so should a functional quantum computer be at hand, they can be easily changed to the wished values). In this demonstration paper, we have therefore frozen the transverse dynamics.

\begin{figure}[h!]
\includegraphics[width=\columnwidth]{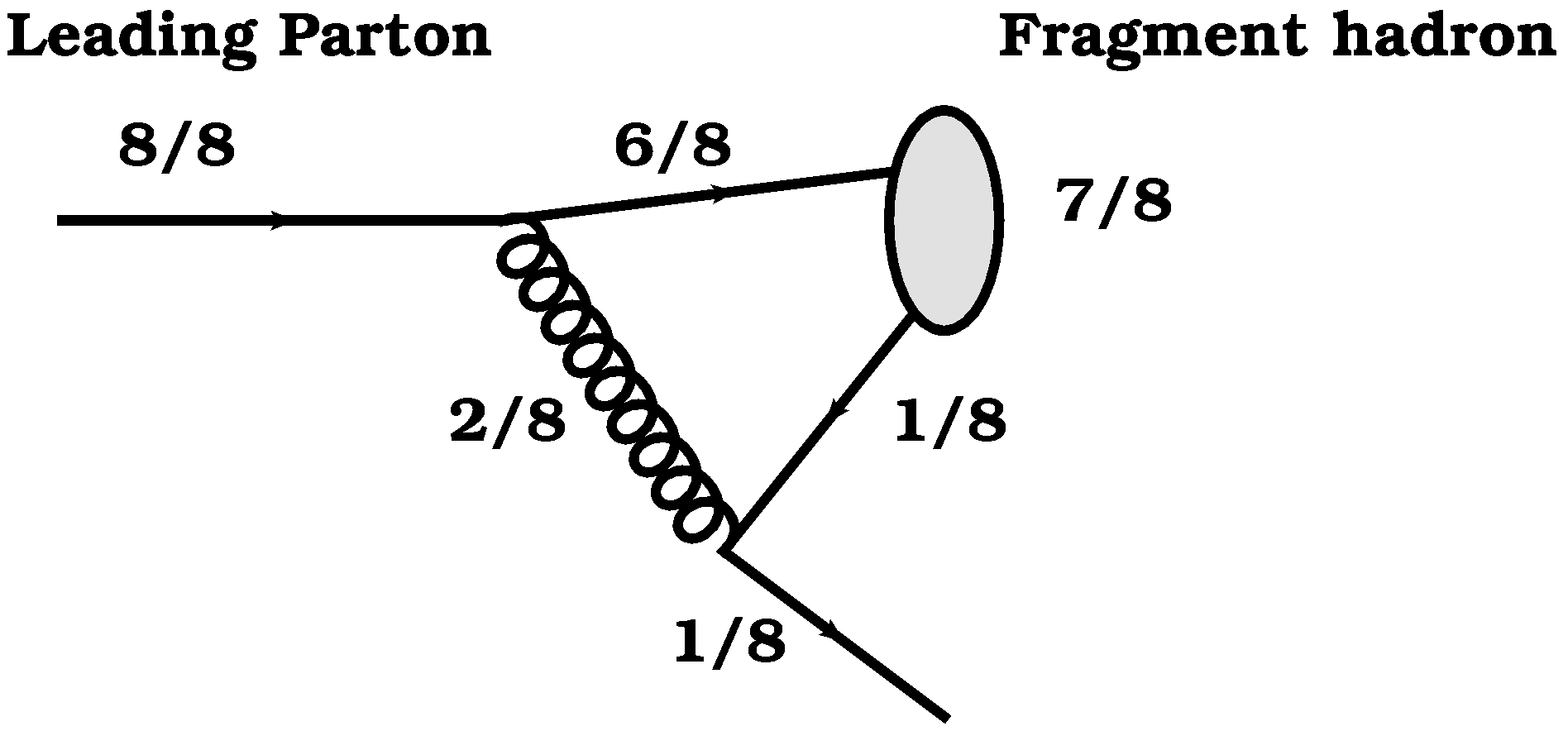}
\caption{Illustration of the maximum momentum carried by the fragmented meson. The initial parton, taken here to be a $c$ quark, must divide its momentum at the first splitting. The gluon then becomes a quark-antiquark pair, each with the minimum momentum representable in the grid ($z=1/8$). This means that the maximum momentum fraction carried by the meson is $z=7/8$. \label{fig:maxz}}
\end{figure}

Since $p_h^+ = \frac{i}{N} p_j^+$, $i=1,2\dots N$, the momentum-fraction variable takes the values $1,2\dots N$.
Therefore, the \emph{minimum} value of a parton's $x$ that we can represent is $1/N$,  an for a hadron with $m$ constituents, the minimum of $z$ is $m/N$. In turn, the \emph{maximum} value, with $m_X$ partons accompanying the fragmented hadron in $|h\ X\rangle$, as depicted in figure~\ref{fig:maxz}, is $1-m_X/N$.

Thus, by definition of the grid, we currently must have zeroes at the endpoints. Improvements over this kinematical setup will require, for example, setting a grid which can represent $p^+=0$ momenta.

%%%%%%%%%%%%%%%%%%%%%%%%%%%%%%%%%%%%%%%%%%%%%%%%%%%%%%%%%%%%%%%%%%%%
\section{Scaling of quantum computing cost}
\label{sec:scaling}
\begin{figure}
    \centering
    \includegraphics[width=0.7\linewidth]{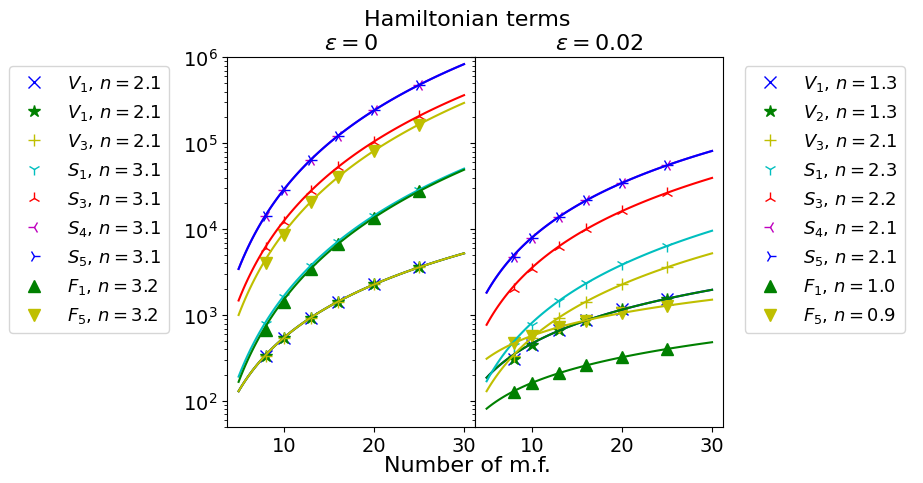}    \caption{Scaling of the number of Hamiltonian terms of each type (Potential $V$, Fork $F$, Seagull $S$)  with the number of encoded momentum fractions (m.f.), fitting the $n$ exponent of Eq.~(\ref{polynomialscaling}). 
    The right plot only includes interactions with $|V|\geq\epsilon|V_{max}|$, with  $\epsilon = 0.02$, to save computational time from terms of size below the noise level. The dominating polynomial-scaling exponents $n$ are indicated. The $\epsilon$ size cutoff barely affects the term $V_3$ whose scaling with $N$ is not modified, because it tends to be numerically large. For other terms there can be substantial gains in execution times.
    }
    \label{fig:SeparateScaling}
\end{figure}

\begin{figure}
    \centering
    \includegraphics[width=0.7\linewidth]{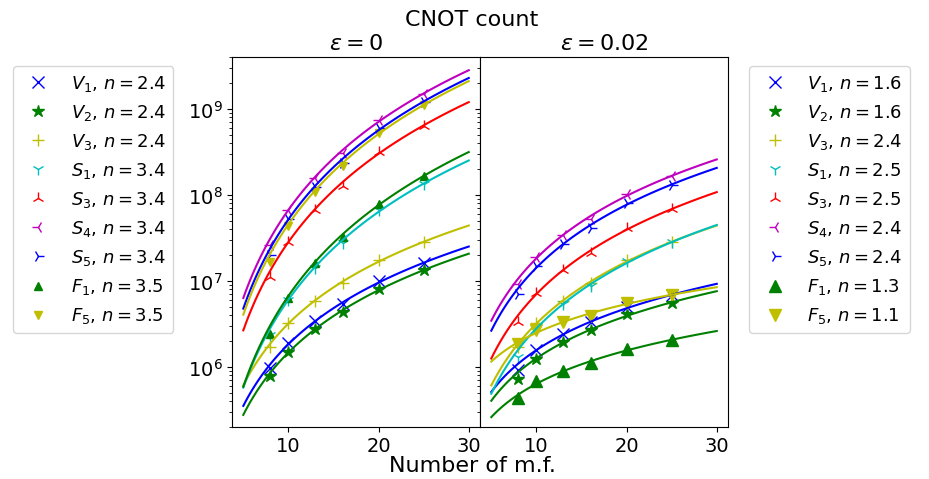}
    \caption{Scaling with the number of encoded momentum fractions (m.f.)  of the number of CNOT gates needed to account for all the possible momentum modes affecting each of the indicated terms of the Hamiltonian. The right plot forces an infrared cutoff to the potential matrix elements, with $|V|\geq\epsilon|V_{max}|$, with $\epsilon = 0.02$, while the computation of the left plot is conducted with $\epsilon=0$, and thus is only cutoff by the grid's volume. The exponents $n$ of the polynomial--scaling law are also given in the legends. The $\epsilon$ size cutoff barely affects the term $V_3$ whose scaling with $N$ is not modified.}
    \label{fig:SeparateCNOTScaling}
\end{figure}

%%%%%%%%%%%%%%%%%%%%%%%%%%%%%%%%%%%%%%%%%%%%%%%%%%%%%%%%%%%%%%%%%%%%
    We now discuss the scaling of resources with the number of longitudinal momentum fractions in $SU(2)$. Tables \ref{tab:HScaling} and \ref{tab:CNOTScaling} give the number of configurations with nonvanishing contribution different from 0 for each Hamiltonian term (ranging from $10^2$ to $10^5$) and the CNOT count (of order $10^3$) per configuration. In turn, Figs.~\ref{fig:SeparateScaling}, \ref{fig:SeparateCNOTScaling} include more points and a fit with the function 
    \begin{equation} \label{polynomialscaling}
     N_{\tt CNOT} =  a\ N^n\ ,
    \end{equation}
    where $a$ and $n$ are fitting parameters and $N$ is the number of available momentum fractions in the partition of the longitudinal momentum  (see table~\ref{tab:parameters_specific}). 
    In calculating the number of gates we note that the decomposition of a controlled gate involving $\nu$ qubits requires $2\nu$ Toffoli gates and therefore $12\nu$ CNOT gates.
    
    We also provide the sensitivity to an infrared cutoff on the potential matrix elements. Given the maximum of the absolute values of all implemented interaction terms, $V_{max}$, we kept only those interactions with a nonnegligible fraction of that maximum value, those for which $|V|\geq \epsilon V_{max}$. The purpose of exploring this restriction is to eliminate a large number of configurations of the various Hamiltonian terms which fall below the noise level in the simulator, and thus, without really contributing to the result of the computation, are consuming resources to basically evaluate  noise. With better hardware such cutoff could become unnecessary. At any rate, because this is a modest-sized exploratory calculation, we examine the systematics by working with and without that cutoff.

\begin{figure}
    \centering
    \includegraphics[width=0.60\linewidth]{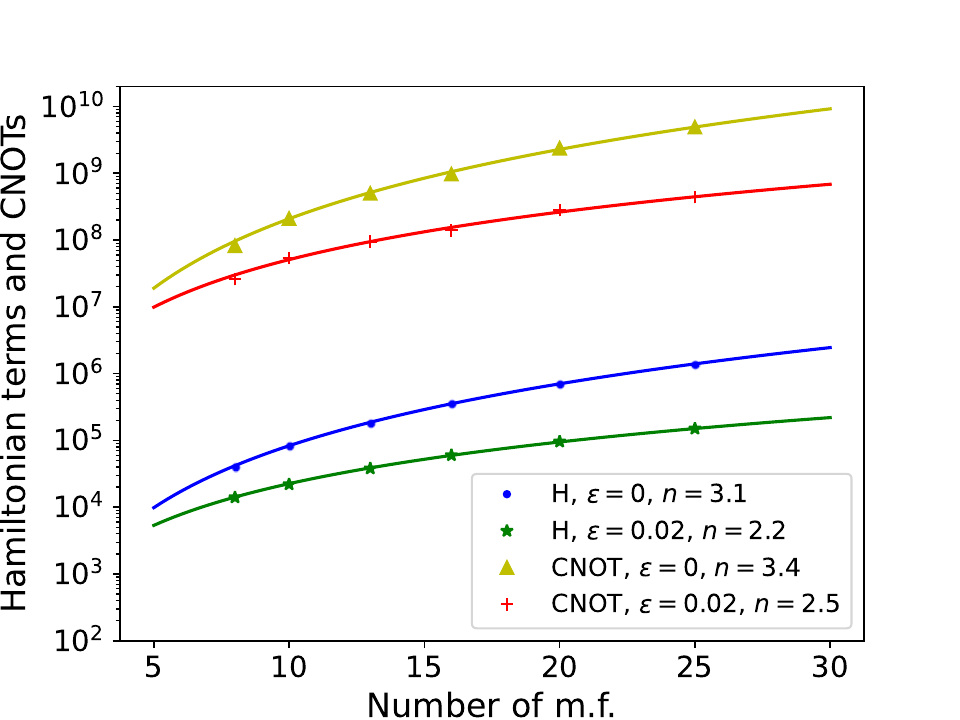}
    \caption{Summary scaling of the number of Hamiltonian terms (lower two lines) and total count of CNOT gates (upper two lines) with the number of encoded $z$ momentum fractions (m.f.). We also show the sensitivity to whether a cutoff on the Hamiltonian matrix element module  $|V|\geq\epsilon|V_{max}|$ is (or is not, by setting $\epsilon=0$) applied. The scaling is polynomial with the exponent $n$ indicated in the legend.}
    \label{fig:totalScaling}
\end{figure}

    Fig.~\ref{fig:totalScaling} summarizes the scaling of all Hamiltonian terms and total CNOT counts.
    
    The terms of the Hamiltonian that show the worst scalings are $S$ and $F$, which involve two--to--two and one--to--three-particle interactions respectively. The somehow smaller costs of the terms of the types $S_1$ and $F_1$ are due to the use of the creation and annihilation operators' anticommutation relations to antisymmetrize the functions in the exponents of the equations analogous to~(\ref{unitaryop}). 
    
    In turn, the largest numbers of CNOT counts of individual quantum-number configurations  are found for the Hamiltonian terms $V_3$ and $F_5$ because they involve the three kinds of particles in our simulation: quarks, antiquarks and gluons. Nevertheless, the total CNOT counts are again largest for the $S$ and $F$ interactions (because in the end, they involve more quantum-number configurations). 
    
    There is an important decrease in resource scaling when imposing the noise-rejecting infrared truncation of the potential via an infrared cutoff, as all of the fit exponents $n$ diminish. From $n\simeq 3.1-3.2$ for the $S$, $F$ terms and $2.1$ for $V$ terms, we find a smaller  $2.3$ for $S$ terms, $1.2$ for $F$ terms and $1.6$ for $V$ terms.
    This is a substantial reduction of the polynomial growth with the number of longitudinal--momentum modes.
    
    An exception is $V_3$, the gluon annihilation into a pair, which takes quite large values,  while the weakest interactions are found in the fork--type vertices.
\begin{table}
    \centering
        \caption{Scaling of the number of different quantum-number configurations for each Hamiltonian piece, with the number of encoded momentum fractions (m.f.). The terms $V_1$, $V_2$ and $V_3$ have the same scaling, as do $S_4$ and $S_5$. In units of $10^3$ configurations.}
    \label{tab:HScaling}
    \begin{tabular}{|c|c|c|c|c|c|c|c|c|c|}
    \hline
    \#m.f. & $V_{1-3}$ & $S_1$ & $S_3$ & $S_4$-$S_5$ & $F_1$ &$F_5$ \\
    \hline
    \hline
     8 &  $0.34$ & $0.78$ & $6.1$ & $14$ & $0.67$ &$4.0$ \\
    \hline    
     16 & $1.4$ & $7.2$ & $53$ & $120$ & $6.7$ &$40$ \\
    \hline
     25 & $3.6$ & $29$ & $210$ & $480$ & $28$ &$170$ \\
    \hline
    \end{tabular}
\end{table}
\begin{table}
    \centering
        \caption{Number of CNOT gates for each quantum-number configuration for the different terms of the Hamiltonian, again scaling with the number of encoded momemtum fractions (m.f.).  In units of $10^3$.}
    \label{tab:CNOTScaling}
    \begin{tabular}{|c|c|c|c|c|c|c|c|c|c|}
    \hline
    \#m.f. & $V_{1}$ & $V_{2}$ &$V_{3}$ & $S_1$ & $S_3$ & $S_{4}$ & $S_{5}$ &  $F_1$ &$F_5$ \\
    \hline
    \hline
     8 &  $2.9$ & $2.3$ & $5.1$ & $3.0$ & $1.8$ & $1.9$ & $1.4$ & $3.6$ &$4.2$ \\
    \hline    
     16 & $3.8$ & $3.0$ & $6.6$ & $3.8$ & $2.4$ & $2.5$ & $2.0$ & $4.8$ &$5.5$ \\
    \hline
     25 & $4.5$ & $3.7$ & $7.9$ & $4.7$ & $3.1$ & $3.2$ & $2.6$ & $6.0$ &$6.7$ \\
    \hline
    \end{tabular}
\end{table}

%%%%%%%%%%%%%%%%%%%%%%%%%%%%%%%%%%%%%%%%%%%%%%%%%%%%%%%%%%%%%%%%%%%%
\section{Systematic uncertainties}\label{app:uncertainty}
%%%%%%%%%%%%%%%%%%%%%%%%%%%%%%%%%%%%%%%%%%%%%%%%%%%%%%%%%%%%%%%%%%%%

There are several sources of uncertainties in the simulation. In Fig.~\ref{fig:JpsifragF} we provided error bands taking into account the currently largest contribution, {\it i.e.}, those corresponding to the time at which fragmentation functions are measured. 

%%%%%%%%%%%%%%%%%%%%%%%%%%%%%%%%%%%%%%%%%%%%%%%%%%%%%%%%%%%%%%%%%%%%
\subsection{Power corrections}
%%%%%%%%%%%%%%%%%%%%%%%%%%%%%%%%%%%%%%%%%%%%%%%%%%%%%%%%%%%%%%%%%%%%
We are not considering possible power corrections of the type $\Lambda_{\rm QCD}/Q$ due to the influence of jet-medium partons on the process (a way to capture some of its effect could also be to modify the simple $J/\psi$ state in Eq.~(\ref{Jpsiwf})). Our computation is limited to the leading-twist fragmentation function $D(z)$ which, by definition, controls the cross-section in Eq.~(\ref{factorization}) and is to be extracted from experiment at asymptotically large $Q$.
Taking into account the drag of the jet medium on the leading parton fragmentation would entail computing higher-twist fragmentation functions. For example, at twist 3, the following matrix element could become relevant, 
$\langle 0 | \psi gF^{+\alpha} |hX\rangle\langle hX \bar{\psi} |0  \rangle$, with an additional gluon field $F$ over the matrix element in Eq.~(\ref{Ddefinition}).
Thus, power corrections do not add uncertainty to $D(z)$ but to the cross-section, which is not our object of study here, by providing additional contributions.

We now discuss three other sources of systematics: Fock-space truncation, Trotterized evolution and the introduction of an infrared cutoff for matrix elements.
%%%%%%%%%%%%%%%%%%%%%%%%%%%%%%%%%%%%%%%%%%%%%%%%%%%%%%%%%%%%%%%%%%%%
\subsection{Uncertainty from truncating  Fock-space particle number}
%%%%%%%%%%%%%%%%%%%%%%%%%%%%%%%%%%%%%%%%%%%%%%%%%%%%%%%%%%%%%%%%%%%%
Each vertex action can emit additional particles (with the Hamiltonian of QCD here used, at most two new ones per insertion), so the
uncertainty from truncating off many--particle states in Fock-space can be seen as limiting the number of vertices (in reality, partially limiting the effect of additional vertices as oscillations between small particle numbers can occur). We exemplify the uncertainty by cutting off the number of insertions of the potential vertices only. The Taylor expansion for the probability of no 1-gluon $q\to qg$ emission in a memory register with capacity for two gluons, up to order $n=5$, reads
\begin{multline}
        P_{2g} \equiv \mel{\Omega_{g_1},\Omega_{g_2} , q_q}{\exp(-iV_1 t)}{\Omega_{g_1},\Omega_{g_2} , q_q}\\\approx1-\frac{(V^2  t^2)}{2}  + \frac{(t^4 \cdot (V^4 + \overline{V}^4))}{4!} +\mathcal{O}(t^6) \label{Prob_2g}
\end{multline}
The potential $ \overline{V^4}\equiv\sum_{j,k} V_{i\rightarrow j}  V_{j\rightarrow k}V_{j\rightarrow k}V_{i\rightarrow j}$, with $i$ the initial quark state,  $j$ and $k$ intermediate states, accessible with successive interactions, under the hypothesis that each interaction lowers the original quark momentum, so that  $p_i>p_j>p_k$. The uncertainty can then be estimated from Taylor's remainder, with $c\in[t_0,t]$, $   R_n(t) = \frac{P_{2g}^{(n+1)}(c)}{(n+1)!} (t - t_0)^{n+1} $ that for $t_0=0$ yields an uncertainty
\begin{equation}
    \Delta P_{2g}\leq  \frac{(V^6+\overline{V}^2\cdot V^4+ \tilde{V} ^6)\cdot t^6}{6!}
\end{equation}
with  $\tilde{V^6}\equiv\sum_{j,k} V_{i\rightarrow j}  V_{j\rightarrow k} V_{k\rightarrow j}V_{j\rightarrow k}V_{k\rightarrow j}  V_{j\rightarrow i}$.

\begin{figure}
    \begin{center}
    \includegraphics[width=0.5\linewidth]{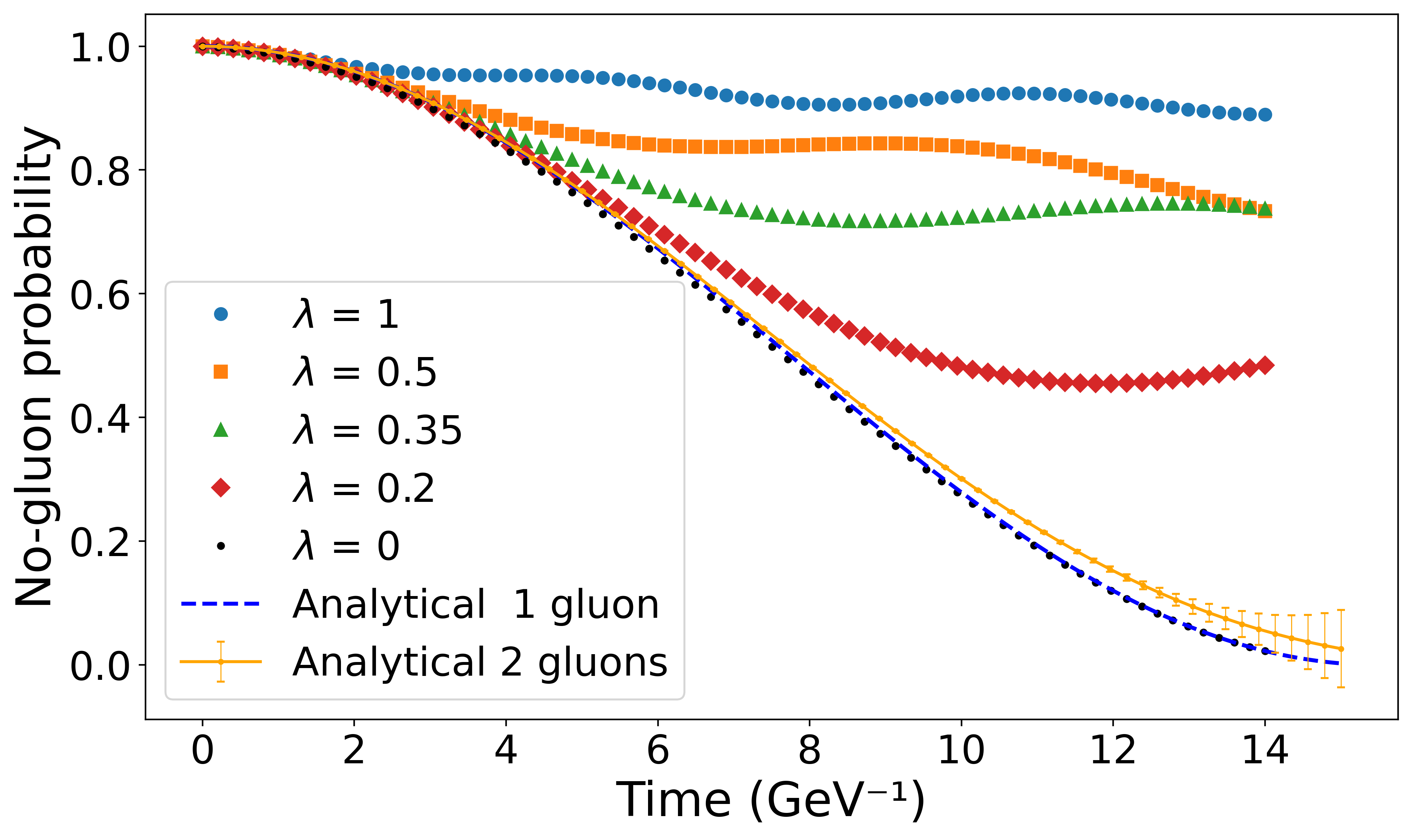}    \caption{Probability that an initial quark with maximum longitudinal momentum is not accompanied by a gluon when it evolves under $\exp\left\{-i\Delta t(V_1+\lambda E_c)\right\}$. Also plotted is the analytically tractable prediction of Eq.~(\ref{Prob_2g}) for the extreme case of $\lambda = 0$. The first minimum of the physical $\lambda = 1$ curve sets the time at which fragmentation functions are measured. This is of interest for the systematic uncertainty induced by the truncation of the Fock space to a finite number of particles.}
    \label{fig:V1+Ec}
    \end{center}
\end{figure}

The theoretical prediction from Eq.~(\ref{Prob_2g}) serves to compare the simulation results using the (unphysical, unless $\lambda=1$) Hamiltonian  $V_1 + \lambda E_c$ with an analytical expression. This is shown in Figure~\ref{fig:V1+Ec}. When $\lambda = 0$ — i.e., only $V_1$ is active—the two curves coincide up to a time given approximately by $14~\text{GeV}^{-1}$, beyond which we would produce a second gluon, but the boundary condition $a^{\dagger}_p\left|g\right\rangle=0$ forced upon us by the limited memory (a second gluon cannot be represented in the classical simulator) begins to introduce deviations between the analytical computation and the simulated one. 

However, for $\lambda\neq 0$ the choice of time to extract the fragmentation function is dominated by the position of the first turning point.
As the kinetic term, initially zero, is gradually activated (increasing $\lambda$ toward 1), the minimum in the no-gluon probability—an effect of finite memory—shifts towards earlier times, as low as $t\sim 4 ~ \text{GeV}^{-1}$. Consequently, the fragmentation function values presented in Fig.~\ref{fig:JpsifragF} are extracted from the simulation around this time.

Simulations employing the vertex interactions (those of order $\mathcal{O}(g)$) and the kinetic energies only, denoted as $V$-simulations,  
are much faster than those in which the terms of $\mathcal{O}(g^2)$ are also taken into account; we denote them by $T$ (from total), see table \ref{tab:parameters_specific} for a summary of running times). The terms of order $\mathcal{O}(g^2)$ (those labelled $S$ and $F$) efficiently spread momentum , so in general the resulting FFs are less intense at high momentum fractions, becoming more concentrated at small $z$.
 
%%%%%%%%%%%%%%%%%%%%%%%%%%%%%%%%%%%%%%%%%%%%%%%%%%%%
\subsection{Uncertainty due to Trotter evolution}
\begin{figure}
    \centering
    \includegraphics[width=0.5\linewidth]{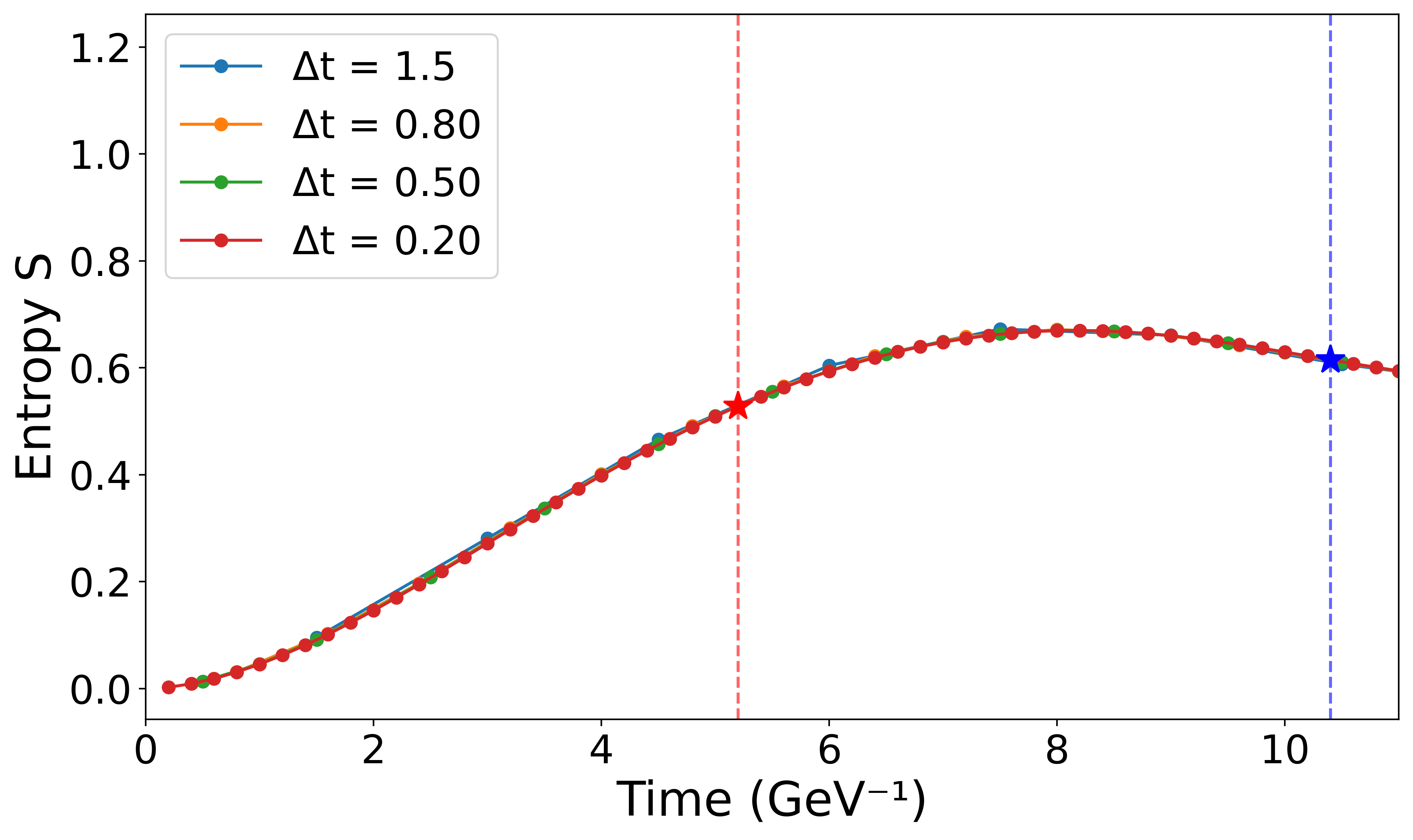}\includegraphics[width=0.5\linewidth]{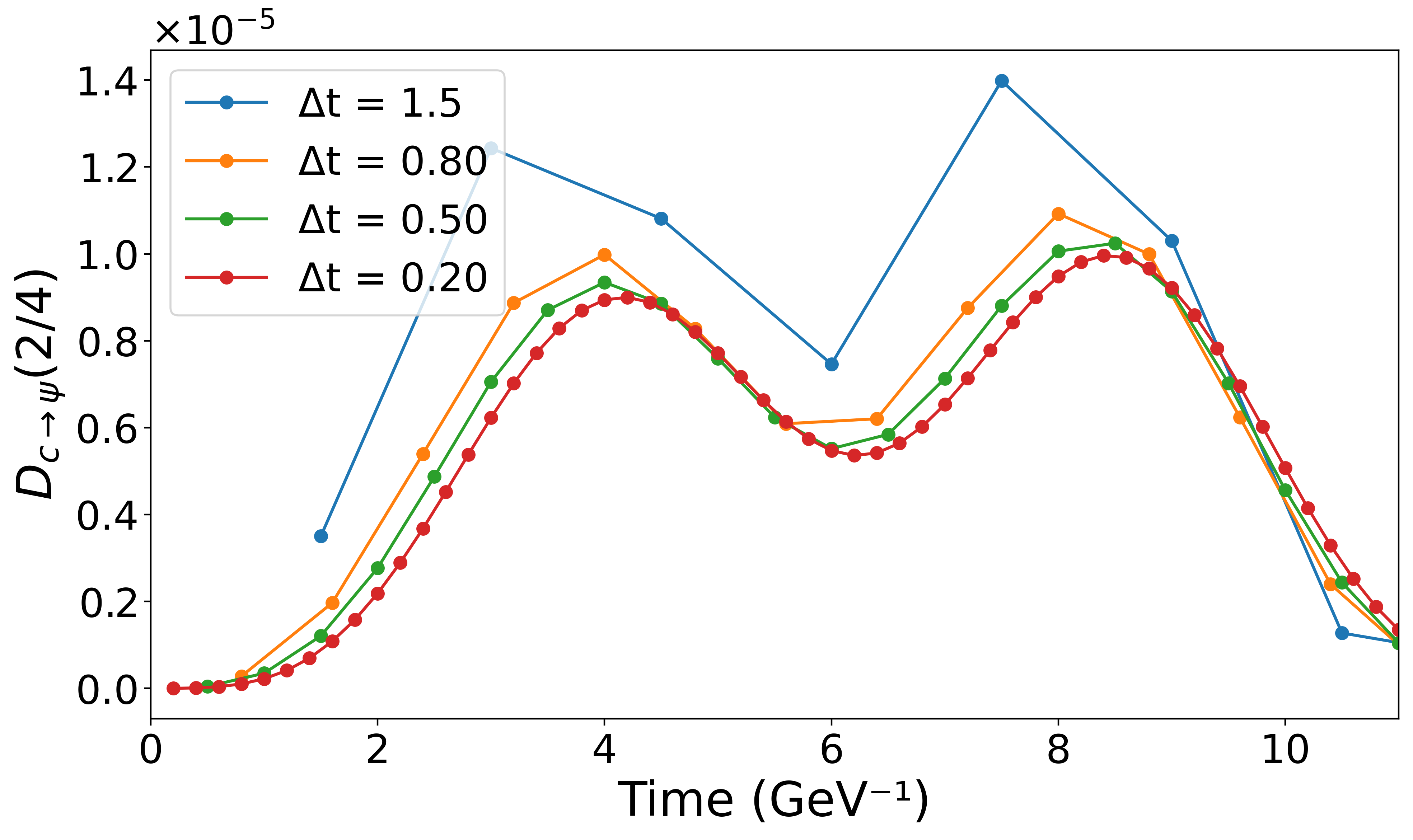}
    \caption{Shannon entropy (left) and fragmentation function at $z=2/4$ (right) as functions of evolution time for different values of the time-step size. Parameters correspond to configuration 1 of table \ref{tab:parameters_specific}. Good convergence with the Trotter time step $\Delta t$ can be appreciated.}
    \label{fig:conf1_stepsize}
\end{figure}

%%%%%%%%%%%%%%%%%%%%%%%%%%%%%%%%%%%%%%%%%%%%%%%%%%%%
The use of the Trotter decomposition to separately exponentiate Hamiltonian terms introduce errors proportional to the square of the time-step size $(\Delta t)^2 $, see table \ref{tab:parameters_specific}. We have calculated the entropy and fragmentation functions for values of the step size down to $\Delta t = 0.1\, \mathrm{GeV}^{-1}$; achieving convergence already at $\Delta t = 0.5\, \mathrm{GeV}^{-1}$ for simulations with kinetic and vertex interactions. See figures ~\ref{fig:conf1_stepsize} and \ref{fig:conf2_epsilon}  for examples of fragmentation functions at $x = 2/4$ and $z = 3/4$ and the entropy for the parameters in configurations 1, 2 and 3 of table \ref{tab:parameters_specific}.

%%%%%%%%%%%%%%%%%%%%%%%%%%%%%%%%%%%%%%%%%%%%%%%%%%%
\subsection{Uncertainty due to infrared cutoff to potential matrix elements}
\begin{figure}
    \centering
    \includegraphics[width=0.5\linewidth]{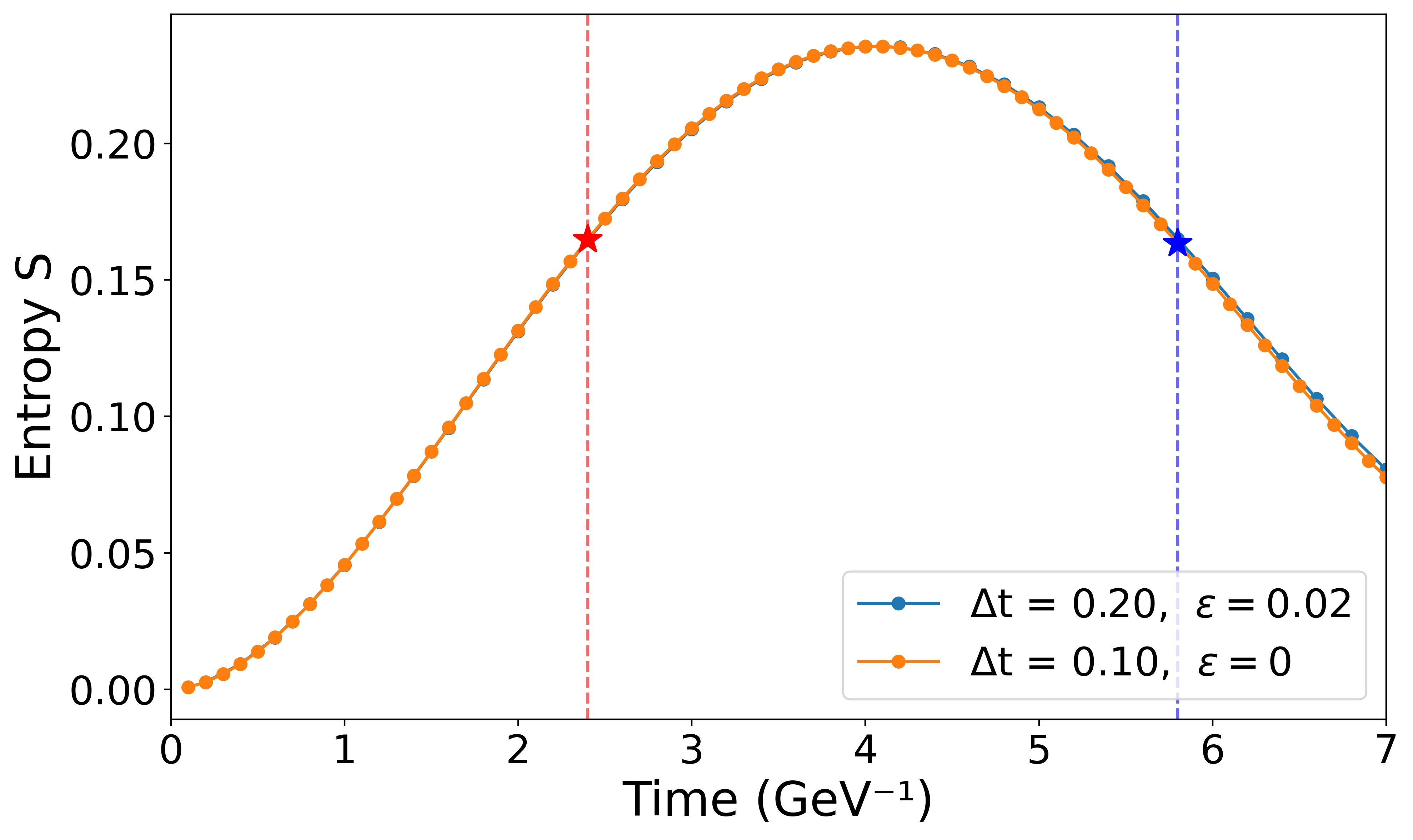}
    \includegraphics[width=0.5\linewidth]{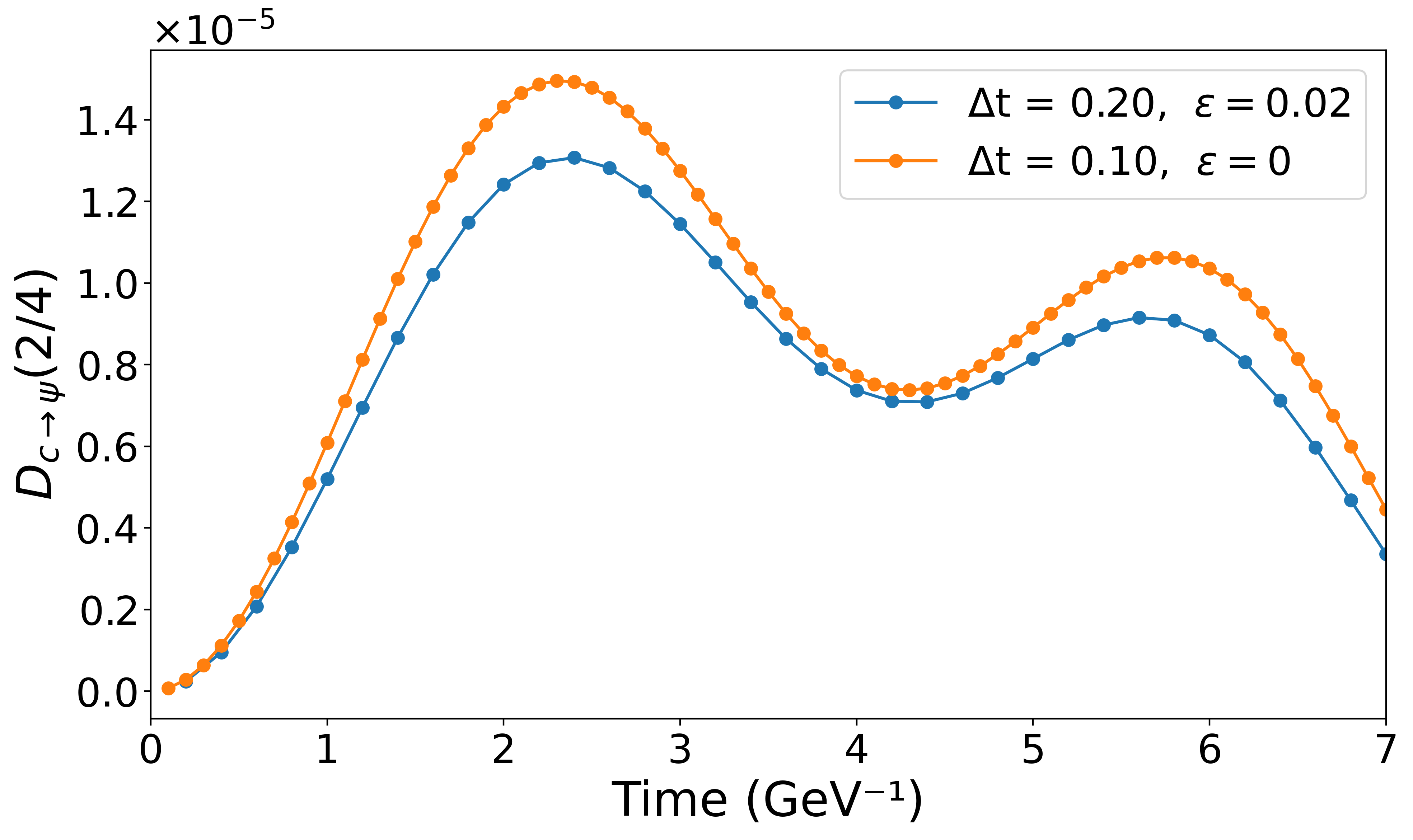}\includegraphics[width=0.5\linewidth]{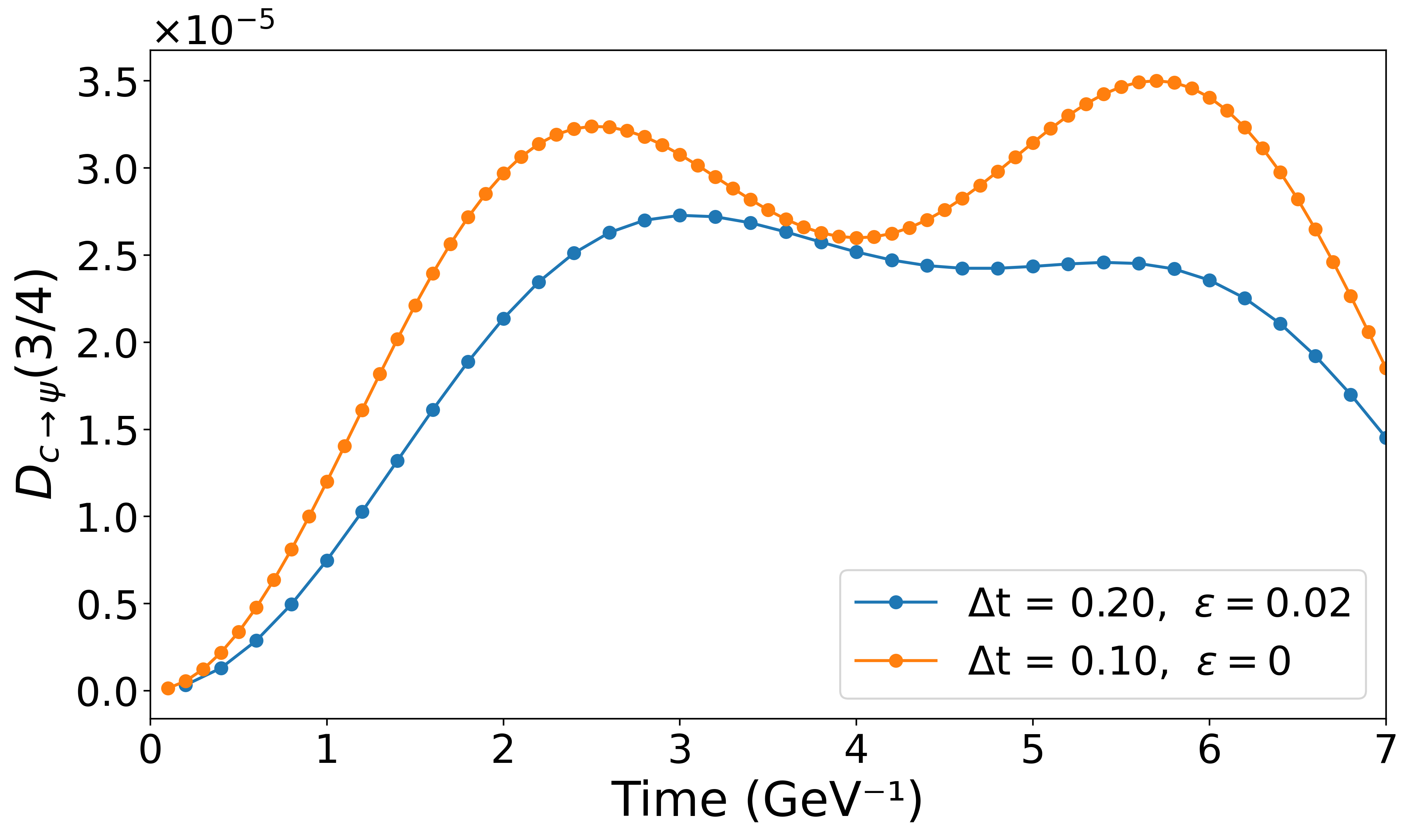}
    \caption{Shannon entropy (upper plot) and fragmentation functions at $z=2/4$ (lower left) and $z=3/4$ (lower right) as functions of evolution time for different values of the Hamiltonian matrix element size cutoff $\epsilon$. Parameters correspond to configurations 2 and 3 of table~\ref{tab:parameters_specific} with $\Delta t$ as indicated in the legends. As can be seen, with our modest number of modes, the fragmentation function still shows oscillations with time which one would expect to be more damped should one have access to a larger quantum memory allowing for further particles and longitudinal momentum partitions.}
    \label{fig:conf2_epsilon}
\end{figure}

%%%%%%%%%%%%%%%%%%%%%%%%%%%%%%%%%%%%%%%%%%%%%%%%%%%%%
Simulations with all terms (T) require exponentiation with interactions mediated by small potentials which we have neglected by the introduction of a cutoff $\epsilon$ as explained in the scaling section \ref{sec:scaling}. The impact of such a parameter is almost negligible in the behavior of the entropy, mostly driven by the change in gluon number, but it can be sizeable on the fragmentation functions, as exemplified in Fig.~\ref{fig:JpsifragF} with the orange and green points and in Fig.~\ref{fig:conf2_epsilon}, in which we compare the cases with $\epsilon=0$ and $\epsilon=0.02$ for a simulation with the $SU(2)$ group and a momentum grid of four points (configurations 2 and 3 of table \ref{tab:parameters_specific}), we see that, in general, the effect of a finite $\epsilon$ is to damp oscillations and decrease the fragmentation functions. The only simulation with an $\epsilon\neq 0$ correspond to configurations $3$ and $5$ 
in table \ref{tab:parameters_specific}. This source of error, combined with the general changes induced by the order $g^2$ terms commented earlier, can explain the drastic decrease of the fragmentation function show in Fig.~\ref{fig:JpsifragF} when all possible Hamiltonian terms within our Fock-space truncation are taken into account.

%%%%%%%%%%%%%%%%%%%%%%%%%%%%%%%%%%%%
\subsection{Uncertainty due to scale choice}
%%%%%%%%%%%%%%%%%%%%%%%%%%%%%%%%%%%%

In a comparison with experiment, one should carefully analyze the scale $\mu$ at which $\alpha_s(\mu)$ is evaluated (in principle, in a full nonperturbative calculation this should be irrelevant, but any grid truncation will introduce a $\mu$--dependent uncertainty). We have opted for fixing the scale $\mu$, and thus $\alpha_s$ to a constant value because our comparison point for the fragmentation function is the NRQCD computation (see Fig~\ref{fig:JpsifragF} in the main text body, with fixed $\alpha_s$). Thus, a fair assessment of this quantum-computer inspired calculation against that template can be achieved even without varying $\mu$.

%%%%%%%%%%%%%%%%%%%%%%%%%%%%%%%%%%%%%%%%%
\subsection{Quality of the wavefunction choice}
%%%%%%%%%%%%%%%%%%%%%%%%%%%%%%%%%%%%%%%%%

In this subsection we confront the uncertainty induced by the simple ansatz for the $J/\psi$ wavefunction, as opposed to having diagonalized the Hamiltonian to obtain it {\it ab initio}. With good quantum hardware this approximation would be unnecessary, as it would be rather straightforward to obtain the ground state in such a ``simple'' channel as the lowest-energy vector charmonium is. Still, in this article with an approximate computation due to our classical simulation of the quantum computer, it is worth ascertaining how close the state is to the actual eigenstate.

To assess it, we have computed 
\begin{eqnarray}
    |\langle J/\psi | U(t) | J/\psi\rangle|^2 &=& \langle U^\dagger \rangle \langle U\rangle \nonumber \\  
    &\simeq& \langle \mathbb{1}+iHt-H^2t^2/2\rangle \langle
        \mathbb{1}-iHt-H^2t^2/2
    \rangle \nonumber \\
 &=& 1-t^2(\langle H \rangle^2 - \langle H^2 \rangle)  \nonumber \\  &=& 1 - t^2 \Delta E^2 \label{UncertaintyH}
\end{eqnarray}
A small-$t$ expansion of the operator then gives us access to the squared energy uncertainty $a=-(\Delta E)^2$
which should vanish in an eigenstate of the Hamiltonian. Thus, its value is a measure of the quality of the wavefunction.

\begin{figure}
    \centering
    \includegraphics[width=\linewidth]{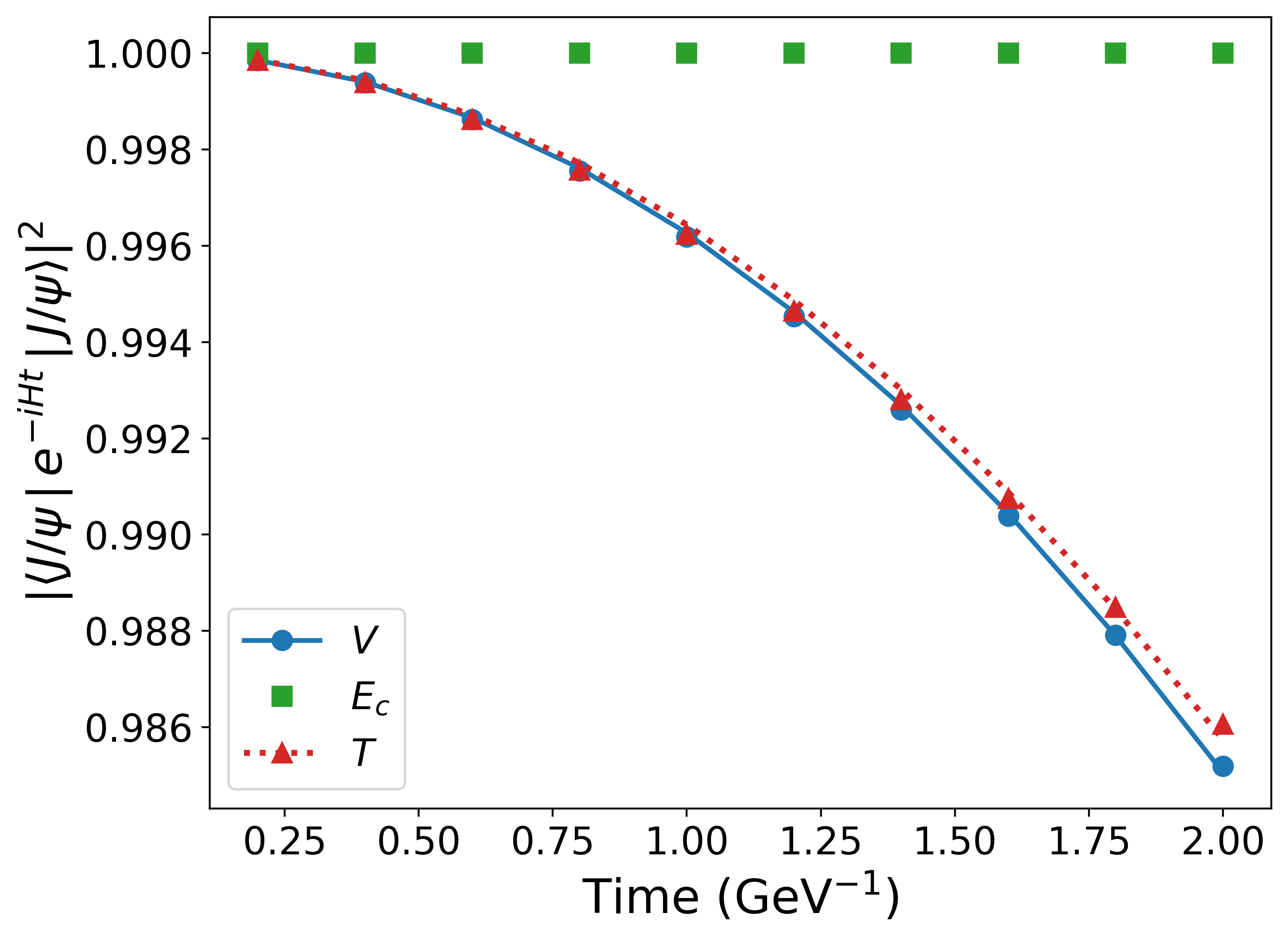}
    \caption{ {\bf Data points}: plot of  $|\langle U\rangle |^2$ in the $J/\psi$ variational state of Eq.~(\ref{Jpsiwf}) from the simulation, for various pieces of the Hamiltonian (free only: $E_c$, $V$-vertices only: $V$, total Hamiltonian, $T$. 
    The calculation is conducted for the $SU(2)$ group and a grid of 4 momentum points.  The quark and antiquark momenta add up to $z = 3/5$.
     {\bf Lines}:
    Quadratic fit as per Eq.~(\ref{UncertaintyH}), yielding $\Delta E = \sqrt{-a} \simeq 61$ MeV. 
    }
    \label{fig:JpsiEuncertainty}
\end{figure}

The dots in Fig.~(\ref{fig:JpsiEuncertainty}) show the value of $\langle U\rangle^2$ for a small $N=4$ longitudinal-momentum grid, with a meson carrying a total momentum fraction of $z = 3/5$. The lines correspond to a parabolic fit of the form
 \begin{equation}
 f(t) = 1+ a  t^2 
 \end{equation}
 whence   $\Delta E$ can be extracted, to each of the data points computed with the kinetic energies only (marked as $E_c$), adding the vertices proportional to $g$ (denoted as $V$)
 and with the total number of terms (shown as $T$).
 Table~\ref{tab:uncertainty} compiles different values of the uncertainty obtained from the various pieces of the Hamiltonian, for values of $z$ ranging from $z=2/4$ to $z =5/4$. Note that it is not possible to generate mesons with total momentum-fractions $z=4/4,5/4$ from an initial parton with momentum fraction $z=4/4$, see subsection (\ref{subsection:momentumgrid}).

 The uncertainty extracted for the kinetic energy alone is close to vanishing, so the proposed discretized wavefunction must be close to an eigenstate of the free Hamiltonian.
There is not much of a difference between including the complete Hamiltonian or just the vertex terms proportional to $g$. 
The resulting uncertainty, depending on $z$, is of order 100 MeV or less, quite smaller than the 3097 MeV of the $J/\psi$ total energy, or the few hundred MeV of its binding energy, and shows that such variational wavefunction is a reasonable starting point to prepare the ground for future machines.

\begin{table}[htbp]
\centering
\caption{Fitted values of the parameter $a = -\Delta E^2$ for different values of $z$ and Hamiltonian terms. \label{tab:uncertainty}}
\resizebox{1\columnwidth}{!}{%
\label{tab:a_values_vs_X}
\begin{tabular}{|c| c |c| c|}
\hline
$z$ & $a_V\left(\mathrm{GeV}^2\right)$ & $a_{E_c} \left(\mathrm{GeV}^2\right)$ & $a_T\left(\mathrm{GeV}^2\right)$ \\
\hline
\hline

2/4 & $-8.857(4)\cdot 10^{-7}$ & --- & $-8.853(4)\cdot 10^{-7}$ \\
\hline

3/4 & $-3.735(11)\cdot 10^{-3}$ & --- & $-3.56(3)\cdot 10^{-3}$ \\
\hline

4/4 & $-1.713(5)\cdot 10^{-2}$ & $-1.097(1)\cdot 10^{-2}$ & $-1.021(6)\cdot 10^{-2}$ \\
\hline

5/4 & $-1.736(6)\cdot 10^{-2}$ & $-1.026(2)\cdot 10^{-2}$ & $-9.42(8)\cdot 10^{-3}$ \\
\hline
\end{tabular}
}
\end{table}

%%%%%%%%%%%%%%%%%%%%%%%%%%%%%%%%%%%%%%%%%
\section{Numerical values of the extracted fragmentation function}
%%%%%%%%%%%%%%%%%%%%%%%%%%%%%%%%%%%%%%%%%

Also in this appendix we wish to list the numerical values and uncertainties of the Fragmentation Function
$D_c^\psi(z)$ presented in Fig.~\ref{fig:JpsifragF} of the main text. These values are given, for $SU(2)$ and $SU(3)$, for the restricted Hamiltonian $V$ and the total one $T$, and for two grids with $N=4$ and $N=8$ longitudinal momentum fraction sizes in table~\ref{tab:fragmentationextraction}. The difference between the two groups stems from larger color factors while holding the coupling constant the same.
It is clear that a working quantum computer will have immense potential in extracting hadron structure functions.

\onecolumngrid


\begin{table}
\caption{Numerical values of the final extraction of the fragmentation function, Fig~\ref{fig:JpsifragF} of the main text, from a classical simulation with the same coding deployable on a quantum computer. In units of $10^{-6}$. The longitudinal momentum--fraction grid has $N=8$ nodes for the first column, $N=4$ in the rest. The $V$ extractions employed a reduced Hamiltonian with only the vertex terms, $T$ is an extraction with the total Hamiltonian. \label{tab:fragmentationextraction}}
\centering
\begin{tabular}{|c|c|c|c|c|c|c|}
\hline
$z$ & SU(2)  & SU(2), T             & SU(2), V & SU(2), T & SU(2), T  & SU(3), V \\
    & V, N=8 & $\epsilon=0.02$, N=8 & & & $\epsilon=0.02$ &  \\
\hline
\hline

2/8 &  $1.1 \pm 0.5$ & $4.4\pm3.0$& -- & -- & -- & -- \\
\hline

3/8 & $1.6 \pm 1.3$ & $1.2\pm0.5$&-- & -- & -- & -- \\
\hline

4/8 & $2.2 \pm 0.7$ &$0.89\pm0.50$& $7.7 \pm 2.3$ & $9.5 \pm 2.1
$ &$ 8.8 \pm 1.7$ & $125 \pm 40$ \\
\hline

5/8 & $4.6 \pm 0.4$ &$1.1\pm0.6$& -- & -- &-- &  -- \\
\hline

6/8 & $8.3 \pm 1.4$ &$1.6\pm0.9$& $42 \pm 18$ & $30 \pm 4$ &$ 25 \pm 2$  & $60 \pm 30$ \\
\hline

7/8 & $11 \pm 4$ &$3.0\pm0.8$& -- & -- & --& -- \\
\hline
\end{tabular}

\end{table}

%\end{document}

\end{document}